\documentclass[10pt]{article}

\usepackage[latin1]{inputenc}
\usepackage{latexsym}
\usepackage{amssymb}
\usepackage{a4}
\usepackage{epsfig,graphics,graphicx}
\usepackage{tabularx}
\usepackage{amsfonts}
\usepackage{amsmath}
\usepackage{pdflscape}
\usepackage{cases}

\usepackage{float}
\usepackage[dvipsnames]{xcolor}
\usepackage{colortbl}
\usepackage{color}
\usepackage{bm}

\usepackage[a4paper,top=2cm,bottom=2cm,left=1.5cm,right=1.5cm]{geometry}
\usepackage{graphicx}
\usepackage{rotating}

\newcommand{\footmsg}[1]{%
  \let\temp\thempfn%
  \def\thempfs{}
  \footnotetext{#1}
  \let\tempfn\temp}

\usepackage{tikz}

\usepackage{enumitem}

\usepackage[at]{easylist}

\usepackage{lipsum,graphicx,subcaption}
\usepackage{dutchcal}

\usepackage{xfrac}
\usepackage{nicefrac}

\usepackage{fancyhdr}            
\fancypagestyle{plain}{%
  \fancyhead{}                                     
  \fancyfoot[CE,CO]{\thepage}       
  \fancyhead[CE,CO]{\textcolor[rgb]{0.64,0.15,0.15}{Published in \emph{International Journal of Solids and Structures} (2019) \textbf{176-177}, 19-35
  \\
doi: https://doi.org/10.1016/j.ijsolstr.2019.07.009}}
\setlength{\headheight}{14.5pt}}

\begin{document}

\newcommand{\beq}{\begin{equation}}
\newcommand{\eeq}{\end{equation}}
\newcommand{\lb}{\label}
\newcommand{\ph}{\phantom}
\newcommand{\beqar}{\begin{eqnarray}}
\newcommand{\eeqar}{\end{eqnarray}}
\newcommand{\barr}{\begin{array}}
\newcommand{\earr}{\end{array}}
\newcommand{\jump}{\parallel}
\newcommand{\Ehat}{\hat{E}}
\newcommand{\That}{\hat{\bf T}}
\newcommand{\Ahat}{\hat{A}}
\newcommand{\chat}{\hat{c}}
\newcommand{\shat}{\hat{s}}
\newcommand{\khat}{\hat{k}}
\newcommand{\muhat}{\hat{\mu}}
\newcommand{\mc}{M^{\scriptscriptstyle C}}
\newcommand{\mei}{M^{\scriptscriptstyle M,EI}}
\newcommand{\mec}{M^{\scriptscriptstyle M,EC}}
\newcommand{\hbeta}{{\hat{\beta}}}
\newcommand{\rec}[2]{\left( #1 #2 \ds{\frac{1}{#1}}\right)}
\newcommand{\rep}[2]{\left( {#1}^2 #2 \ds{\frac{1}{{#1}^2}}\right)}
\newcommand{\derp}[2]{\ds{\frac {\partial #1}{\partial #2}}}
\newcommand{\derpn}[3]{\ds{\frac {\partial^{#3}#1}{\partial #2^{#3}}}}
\newcommand{\dert}[2]{\ds{\frac {d #1}{d #2}}}
\newcommand{\dertn}[3]{\ds{\frac {d^{#3} #1}{d #2^{#3}}}}

\def\c{{\circ}}
\def\bob{{\, \underline{\overline{\otimes}} \,}}
\def\ob{{\, \underline{\otimes} \,}}
\def\scalp{\mbox{\boldmath$\, \cdot \, $}}
\def\scalpp{\mbox{\boldmath$\, : \, $}}
\def\gdp{\makebox{\raisebox{-.215ex}{$\Box$}\hspace{-.778em}$\times$}}
\def\daa{\makebox{\raisebox{-.050ex}{$-$}\hspace{-.550em}$: ~$}}
\def\mK{\mbox{${\mathcal{K}}$}}
\def\cK{\mbox{${\mathbb {K}}$}}

\def\Xint#1{\mathchoice
   {\XXint\displaystyle\textstyle{#1}}%
   {\XXint\textstyle\scriptstyle{#1}}%
   {\XXint\scriptstyle\scriptscriptstyle{#1}}%
   {\XXint\scriptscriptstyle\scriptscriptstyle{#1}}%
   \!\int}
\def\XXint#1#2#3{{\setbox0=\hbox{$#1{#2#3}{\int}$}
     \vcenter{\hbox{$#2#3$}}\kern-.5\wd0}}
\def\ddashint{\Xint=}
\def\fpint{\Xint=}
\def\dashint{\Xint-}
\def\cpvint{\Xint-}
\def\intl{\int\limits}
\def\cpvintl{\cpvint\limits}
\def\fpintl{\fpint\limits}
\def\ointl{\oint\limits}
\def\bA{{\bf A}}
\def\bB{{\bf B}}
\def\bC{{\bf C}}
\def\bD{{\bf D}}
\def\bE{{\bf E}}
\def\bF{{\bf F}}
\def\bG{{\bf G}}
\def\bH{{\bf H}}
\def\bI{{\bf I}}
\def\bJ{{\bf J}}
\def\bK{{\bf K}}
\def\bL{{\bf L}}
\def\bM{{\bf M}}
\def\bN{{\bf N}}
\def\bO{{\bf O}}
\def\b0{{\bf 0}}
\def\bP{{\bf P}}
\def\bQ{{\bf Q}}
\def\bR{{\bf R}}
\def\bS{{\bf S}}
\def\bT{{\bf T}}
\def\bU{{\bf U}}
\def\bV{{\bf V}}
\def\bW{{\bf W}}
\def\bX{{\bf X}}
\def\bY{{\bf Y}}
\def\bZ{{\bf Z}}

\def\ba{{\bf a}}
\def\bb{{\bf b}}
\def\bc{{\bf c}}
\def\bd{{\bf d}}
\def\be{{\bf e}}
\def\bbf{{\bf f}}
\def\bg{{\bf g}}
\def\bh{{\bf h}}
\def\bi{{\bf i}}
\def\bj{{\bf j}}
\def\bk{{\bf k}}
\def\bl{{\bf l}}
\def\bm{{\bf m}}
\def\bn{{\bf n}}
\def\bo{{\bf o}}
\def\bp{{\bf p}}
\def\bq{{\bf q}}
\def\br{{\bf r}}
\def\bs{{\bf s}}
\def\bt{{\bf t}}
\def\bu{{\bf u}}
\def\bv{{\bf v}}
\def\bw{{\bf w}}
\def\bx{{\bf x}}
\def\by{{\bf y}}
\def\bz{{\bf z}}

\def\bxi{\mbox{\boldmath${\xi}$}}
\def\balpha{\mbox{\boldmath${\alpha}$}}
\def\bbeta{\mbox{\boldmath${\beta}$}}
\def\bgamma{\mbox{\boldmath${\gamma}$}}
\def\bepsilon{\mbox{\boldmath${\epsilon}$}}
\def\bvarepsilon{\mbox{\boldmath${\varepsilon}$}}
\def\bomega{\mbox{\boldmath${\omega}$}}
\def\bphi{\mbox{\boldmath${\phi}$}}
\def\bsigma{\mbox{\boldmath${\sigma}$}}
\def\bfeta{\mbox{\boldmath${\eta}$}}
\def\bDelta{\mbox{\boldmath${\Delta}$}}
\def\bdelta{\mbox{\boldmath${\delta}$}}
\def\btau{\mbox{\boldmath $\tau$}}
\def\bmu{\mbox{\boldmath $\mu$}}
\def\bchi{\mbox{\boldmath $\chi$}}
\def\bnabla{\mbox{\boldmath $\nabla$}}
\def\tr{{\rm tr}}
\def\dev{{\rm dev}}
\def\div{{\rm div}}
\def\Div{{\rm Div}}
\def\Grad{{\rm Grad}}
\def\grad{{\rm grad}}
\def\Lin{{\rm Lin}}
\def\Sym{{\rm Sym}}
\def\Skw{{\rm Skew}}
\def\abs{{\rm abs}}
\def\Re{{\rm Re}}
\def\Im{{\rm Im}}
\def\capB{\mbox{\boldmath${\mathsf B}$}}
\def\capC{\mbox{\boldmath${\mathsf C}$}}
\def\capD{\mbox{\boldmath${\mathsf D}$}}
\def\capE{\mbox{\boldmath${\mathsf E}$}}
\def\capG{\mbox{\boldmath${\mathsf G}$}}
\def\tcapG{\tilde{\capG}}
\def\capH{\mbox{\boldmath${\mathsf H}$}}
\def\capK{\mbox{\boldmath${\mathsf K}$}}
\def\capL{\mbox{\boldmath${\mathsf L}$}}
\def\capM{\mbox{\boldmath${\mathsf M}$}}
\def\capR{\mbox{\boldmath${\mathsf R}$}}
\def\capW{\mbox{\boldmath${\mathsf W}$}}

\def\i{\mbox{${\mathrm i}$}}
\def\mC{\mbox{\boldmath${\mathcal C}$}}
\def\mB{\mbox{${\mathcal B}$}}
\def\mE{\mbox{${\mathcal{E}}$}}
\def\mL{\mbox{${\mathcal{L}}$}}
\def\mK{\mbox{${\mathcal{K}}$}}
\def\mV{\mbox{${\mathcal{V}}$}}
\def\C{\mbox{\boldmath${\mathcal C}$}}
\def\E{\mbox{\boldmath${\mathcal E}$}}

\def\AAM{{\it Advances in Applied Mechanics }}
\def\ACME{{\it Arch. Comput. Meth. Engng.}}
\def\ARMA{{\it Arch. Rat. Mech. Analysis}}
\def\AMR{{\it Appl. Mech. Rev.}}
\def\ASCEEM{{\it ASCE J. Eng. Mech.}}
\def\ACTA{{\it Acta Mater.}}
\def\CMAME {{\it Comput. Meth. Appl. Mech. Engrg.}}
\def\CRAS{{\it C. R. Acad. Sci. Paris}}
\def\CRM{{\it Comptes Rendus M\'ecanique}}
\def\EFM{{\it Eng. Fracture Mechanics}}
\def\EJMA{{\it Eur.~J.~Mechanics-A/Solids}}
\def\IJES{{\it Int. J. Eng. Sci.}}
\def\IJF{{\it Int. J. Fracture}}
\def\IJMS{{\it Int. J. Mech. Sci.}}
\def\IJNAMG{{\it Int. J. Numer. Anal. Meth. Geomech.}}
\def\IJP{{\it Int. J. Plasticity}}
\def\IJSS{{\it Int. J. Solids Structures}}
\def\IngA{{\it Ing. Archiv}}
\def\JAM{{\it J. Appl. Mech.}}
\def\JAP{{\it J. Appl. Phys.}}
\def\JAE{{\it J. Aerospace Eng.}}
\def\JE{{\it J. Elasticity}}
\def\JM{{\it J. de M\'ecanique}}
\def\JMPS{{\it J. Mech. Phys. Solids}}
\def\JSV{{\it J. Sound and Vibration}}
\def\MACRO{{\it Macromolecules}}
\def\MMT{{\it Mech. Mach. Th.}}
\def\MOM{{\it Mech. Materials}}
\def\MMS{{\it Math. Mech. Solids}}
\def\MMT{{\it Metall. Mater. Trans. A}}
\def\MPCPS{{\it Math. Proc. Camb. Phil. Soc.}}
\def\MSE{{\it Mater. Sci. Eng.}}
\def\NATURE{{\it Nature}}
\def\NATUREM{{\it Nature Mater.}}
\def\PHIL{{\it Phil. Trans. R. Soc.}}
\def\PMPS{{\it Proc. Math. Phys. Soc.}}
\def\PNAS{{\it Proc. Nat. Acad. Sci.}}
\def\PRE{{\it Phys. Rev. E}}
\def\PRL{{\it Phys. Rev. Letters}}
\def\PRSL{{\it Proc. R. Soc.}}
\def\ROCK{{\it Rock Mech. and Rock Eng.}}
\def\QAM{{\it Quart. Appl. Math.}}
\def\QJMAM{{\it Quart. J. Mech. Appl. Math.}}
\def\SCIENCE{{\it Science}}
\def\SCRMAT{{\it Scripta Mater.}}
\def\SM{{\it Scripta Metall.}}
\def\ZAMM{{\it Z. Angew. Math. Mech.}}
\def\ZAMP{{\it Z. Angew. Math. Phys.}}
\def\ZVDI{{\it Z. Verein. Deut. Ing.}}

\title{Identification of second-gradient elastic materials from planar hexagonal lattices. 
Part II: Mechanical characteristics and model validation}

\author{G. Rizzi, F. Dal Corso, D. Veber and D. Bigoni$^1$ \\ 
DICAM, University of Trento\\
via Mesiano 77, I-38123 Trento, Italy
}
\date{}
\maketitle
\footnotetext[1]{Corresponding author: Davide Bigoni 
fax: +39 0461 282599; tel.: +39 0461 282507; web-site:
http://www.ing.unitn.it/$\sim$bigoni/; e-mail:
davide.bigoni@unitn.it.}

\begin{abstract}

Positive definiteness and symmetry of the constitutive tensors describing a second-gradient elastic (SGE) material, which is energetically 
equivalent to a hexagonal planar lattice made up of axially deformable bars, are analyzed by exploiting the closed form-expressions obtained in part I of the present study in the \lq condensed' form. It is shown that, while the first-order approximation leads to an isotropic Cauchy material, a second-order identification procedure provides an equivalent model exhibiting non-locality, non-centrosymmetry, and 
anisotropy. 
The derivation of the constitutive properties for the SGE from those of the \lq condensed' one (obtained by considering a quadratic remote displacement which generates stress states satisfying equilibrium) is presented. 
Comparisons between the mechanical responses of the periodic lattice and of the equivalent SGE material under simple shear and uniaxial strain show the efficacy of the proposed identification procedure and therefore validate the proposed constitutive model. This model reveals that, at higher-order, a lattice material can be made equivalent to a second-gradient elastic material exhibiting an internal length, a finding which is now open for applications in micromechanics.
\end{abstract}

\vspace{10 mm}

\section{Introduction}

Higher-order constitutive equations for microstructured solids \cite{bigonipanos, gourgiotis2018, GOURGIOTIS2016169, papathanasiou2016, zisis2014some, zisis2018}, initially introduced following a phenomenological approach \cite{cosserat,fleckhutchinson, fleckwillis, koiter, mindlin1964micro,MINDLIN1968109}, are nowadays more and more often 
connected to material microstructures by means of various homogenization schemes \cite{bacigalupo2014second3,bacigalupo2010second1,BACIGALUPO2016126,bacigalupo2017wave,dalcorsodeseri,druganwillis,forest1998mechanics, forest, forest2002, smyshlyaev2000, smyshlyaev2009, tran2012, trova, ostoja1999, willis}.

\begin{figure}[H]
	\centering
	\includegraphics[width=\textwidth]{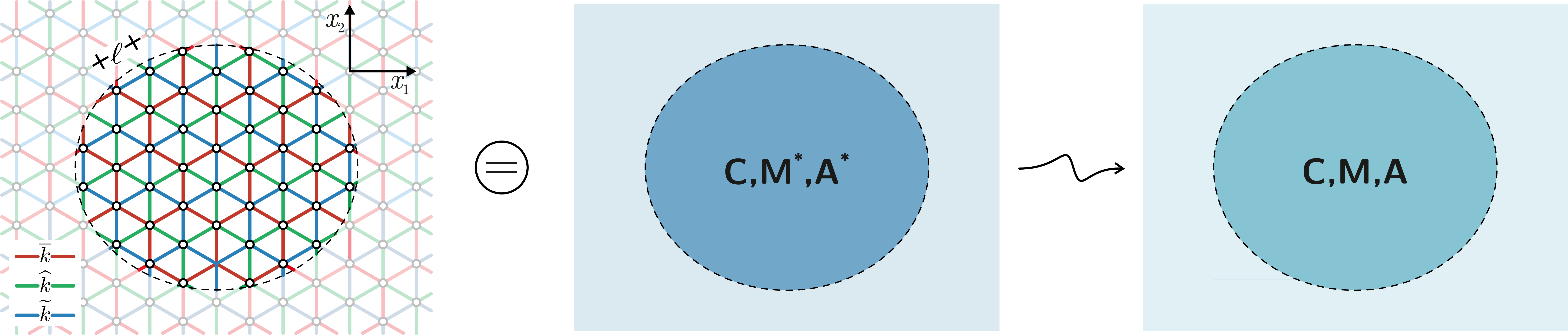}
	\caption{The \lq condensed' second-gradient elastic material (center), characterized by the constitutive matrices $\boldsymbol{\mathsf{C}}$, $\boldsymbol{\mathsf{M}}^*$, and $\boldsymbol{\mathsf{A}}^*$ and identified in Part I of this article as  the solid energetically equivalent to the periodic lattice  (left), which is based on a hexagonal unit cell of side length $\ell$ and is composed of elastic bars, with three different values of axial stiffness $\overline{k}$, $\widetilde{k}$ and $\widehat{k}$ (distinguished  through different colours) and hinged to each other. The second-gradient elastic material (right), characterized by matrices $\boldsymbol{\mathsf{C}}$, $\boldsymbol{\mathsf{M}}$, and $\boldsymbol{\mathsf{A}}$, is  obtained from a relaxation of the \lq condensed' solid.}
	\label{fig:DisCont}
\end{figure}

In the context of deriving the higher-order parameters from discrete microstructures, 
closed-form expressions are obtained  
in part I of the present study \cite{rizzipt1}, 
for the constitutive tensors characterizing a homogeneous second-gradient elastic material ($\mathsf{SGE}$), equivalent to the hexagonal planar lattice 
sketched in Fig. \ref{fig:DisCont}. 
The lattice is based on a unit cell of side length $\ell$ and composed of axially deformable bars with stiffnesses $\overline{k}$, $\widehat{k}$, and $\widetilde{k}$. The
equivalent second-gradient elastic material is obtained by imposing a matching between its strain energy and that of the lattice, when both the solid and the lattice are subject to remote boundary conditions, which realize all possible  
quadratic displacement fields, but constrained by the requirement that the generated stress fields are in equilibrium (without body forces). 
In this way, \lq condensed forms' for the higher-order constitutive tensors follow, because the equilibrium constraint may in a sense be interpreted as a lack in the  number of \lq independent displacement tests' sufficient to completely characterize a second-gradient elastic material. The developed elastic model always reduces to the isotropic Cauchy material defined by \cite{day1992elastic,snyder1992elastic}, when the hexagon side length $\ell$ is set to be equal to zero.

In the present paper, a standard $\mathsf{SGE}$ material (Fig. \ref{fig:DisCont}, right) is derived from the \lq condensed' version obtained in Part I of this study (sketched in Fig. \ref{fig:DisCont}, center). The derivation is pursued through a simple approach, which corresponds to the imposition of the equivalence between the elastic energies of the lattice and of the equivalent material, but neglecting the equilibrium constraint on the displacement fields. Interestingly, such a simple approach preserves the symmetry properties of the `condensed' material and leaves unchanged its positive definiteness domain. Following this approach, the most important result of this paper is obtained, which is provided by three constitutive matrices defining the $\mathsf{SGE}$ material equivalent to the lattice:
\begin{equation}
\begin{array}{cccc}
&\boldsymbol{\mathsf{C}}\left(\overline{k}, \widehat{k}, \widetilde{k}\right)=
\left(
\begin{array}{cccccc}
	\lambda' + 2\mu & \lambda' & 0  \\ [2mm]
	\lambda' & \lambda' + 2\mu & 0  \\ [2mm]
	0 & 0 & \mu  \\ [2mm]
\end{array}\right),
\\[10mm]
&\boldsymbol{\mathsf{M}}\left(\overline{k}, \widehat{k}, \widetilde{k}\right) = 
\mathsf{m}\ell 
\left(
\begin{array}{cccccc}
	0 & 0 & 1 & -1 & 1 & 0 \\
	0 & 0 & 1 & -1 & 1 & 0 \\
	0 & 0 & 0 & 0 & 0 & 0
\end{array}\right),
\\[10mm]
&\boldsymbol{\mathsf{A}}\left(\overline{k}, \widehat{k}, \widetilde{k}\right)=\ell^2
\resizebox{0.82\textwidth}{!}{$
\left(
\begin{array}{cccccc}
	\mathsf{a}_{11} & \mathsf{a}_{12} & 0 & 0 & 0 & \mathsf{a}_{16} \\ [2mm]
	\mathsf{a}_{12} & \mathsf{a}_{11}-2 (\mathsf{a}_{16}+\mathsf{a}_{26}) & 0 & 0 & 0 & \mathsf{a}_{26} \\ [2mm]
	0 & 0 & \mathsf{a}_{11}+\mathsf{a}_{12}-2 (\mathsf{a}_{16}+\mathsf{a}_{26})-\mathsf{a}_{34} & \mathsf{a}_{34} & \mathsf{a}_{12}+\mathsf{a}_{26}-\mathsf{a}_{34} & 0 \\ [2mm]
	0 & 0 & \mathsf{a}_{34} & \mathsf{a}_{11}+\mathsf{a}_{12}-\mathsf{a}_{34} & -\mathsf{a}_{12}+\mathsf{a}_{16}+\mathsf{a}_{34} & 0 \\ [2mm]
	0 & 0 & \mathsf{a}_{12}+\mathsf{a}_{26}-\mathsf{a}_{34} & -\mathsf{a}_{12}+\mathsf{a}_{16}+\mathsf{a}_{34} & \dfrac{\mathsf{a}_{11}+\mathsf{a}_{12}-\mathsf{a}_{16}-\mathsf{a}_{26}-2 \mathsf{a}_{34}}{2}  & 0 \\ [2mm]
	\mathsf{a}_{16} & \mathsf{a}_{26} & 0 & 0 & 0 & \dfrac{\mathsf{a}_{11}-\mathsf{a}_{12}-\mathsf{a}_{16}-\mathsf{a}_{26}}{2} 
\end{array}\right),
$}
\end{array}
\label{eq:effettivo}
\end{equation}
where the constants for the Cauchy part (expressed using the Lamé constants) of the equivalent elastic material are 
\begin{equation}
\begin{array}{lcl}
	\hspace*{0.2cm}
\lambda' = \dfrac{2I_{[1]}I_{[2]}-9I_{[3]}}{4\sqrt{3} I_{[2]}},
\quad \mu = \dfrac{9I_{[3]}}{4\sqrt{3}I_{[2]}},
\end{array}
\label{eq:coeff_exte_C}
\end{equation}
while those for the non centro-symmetric part are 
\begin{equation}
\begin{array}{lcl}
\quad \mathsf{m} = \dfrac{\left(\widehat{k}-\widetilde{k}\right) \left[\widehat{k} \widetilde{k}-2 \overline{k} (\widehat{k}+\widetilde{k})\right]}{8 \sqrt{3} I_{[2]}}, 
\end{array}
\label{eq:coeff_exte_M}
\end{equation}
and those for the curvature part are
\begin{equation}
\begin{array}{lcl}
	\mathsf{a}_{11}= &\dfrac{\sqrt{3}}{576  I_{[2]}^3}& \left[48 \overline{k}^5 \left(\widehat{k}+\widetilde{k}\right)^2+4 \overline{k}^4 \left(\widehat{k}+\widetilde{k}\right) \left(31 \widehat{k}^2+194 \widehat{k} \widetilde{k}+31 \widetilde{k}^2\right)\right.\\
	&&\left. +\overline{k}^3 \left(28 \widehat{k}^4+814 \widehat{k}^3 \widetilde{k}+3024 \widehat{k}^2 \widetilde{k}^2+814 \widehat{k}\widetilde{k}^3+28 \widetilde{k}^4\right)\right.\\
	&&\left.+6 I_{[3]} \overline{k}\left(\widehat{k}+\widetilde{k}\right) \left(14 \widehat{k}^2+263 \widehat{k} \widetilde{k}+14 \widetilde{k}^2\right)\right.\\
	&&\left.+I_{[3]} \widehat{k} \widetilde{k} \left(84 \widehat{k}^2+757 \widehat{k} \widetilde{k}+84 \widetilde{k}^2\right)+28\widehat{k}^3 \widetilde{k}^3 \left(\widehat{k}+\widetilde{k}\right)\right], \\
	\mathsf{a}_{12}=& \dfrac{3 \sqrt{3} I_{[3]} }{64  I_{[2]}^3} & \left[ 8 \overline{k}^3 \left(\widehat{k}+\widetilde{k}\right)+\overline{k}^2 \left(8 \widehat{k}^2+60 \widehat{k} \widetilde{k}+8 \widetilde{k}^2\right)+20 I_{[3]} \left(\widehat{k}+\widetilde{k}\right)+3 \widehat{k}^2 \widetilde{k}^2\right], \\
	\mathsf{a}_{16}=&\dfrac{\sqrt{3}}{576  I_{[2]}^2} & \left[24 \overline{k}^4 \left(\widehat{k}+\widetilde{k}\right)+\overline{k}^3 \left(62 \widehat{k}^2+256 \widehat{k} \widetilde{k}+62 \widetilde{k}^2\right) \right.\\
	&&+\overline{k}^2\left(\widehat{k}+\widetilde{k}\right) \left(14 \widehat{k}^2+299 \widehat{k} \widetilde{k}+14 \widetilde{k}^2\right)\\
	&&\left.+4 I_{[3]} \left(7\widehat{k}^2+19 \widehat{k} \widetilde{k}+7 \widetilde{k}^2\right)+14 \widehat{k}^2 \widetilde{k}^2 \left(\widehat{k}+\widetilde{k}\right)\right], \\
	\mathsf{a}_{26}=&\dfrac{\sqrt{3} I_{[3]} }{64  I_{[2]}^2}&\left[12 \overline{k}^2+7 \overline{k} \left(\widehat{k}+\widetilde{k}\right)+22 \widehat{k} \widetilde{k}\right], \\
	\mathsf{a}_{34}=&\dfrac{\sqrt{3}}{192  I_{[2]}^3}&\left[72 I_{[3]} \overline{k}^3 \left(\widehat{k}+\widetilde{k}\right)+\overline{k}^3 \left(-8 \widehat{k}^4+40 \widehat{k}^3 \widetilde{k}+492 \widehat{k}^2 \widetilde{k}^2+40 \widehat{k} \widetilde{k}^3-8 \widetilde{k}^4\right)\right.\\
	&&\left.+234 I_{[3]}^{2} \left(\widehat{k}+\widetilde{k}\right)+3 I_{[3]} \widehat{k} \widetilde{k} \left(2 \widehat{k}^2+31 \widehat{k} \widetilde{k}+2 \widetilde{k}^2\right)-2 \widehat{k}^3 \widetilde{k}^3 \left(\widehat{k}+\widetilde{k}\right)\right] .
\end{array}
\label{eq:coeff_exte_A}
\end{equation}
Note that in the above equations $\lambda'=\lambda$ ($\lambda$ being the Lamé constant) applies when \textit{plane strain} prevails (while in the case of plane stress 
$\lambda'=2\lambda\mu/(\lambda + 2\mu)$) 
and
\begin{equation}
I_{[1]}=\overline{k} + \widehat{k} + \widetilde{k},\qquad I_{[2]}=\overline{k}\,\widehat{k} + \overline{k}\,\widetilde{k} + \widetilde{k}\,\widehat{k},\qquad I_{[3]}=\overline{k}\,\widehat{k}\,\widetilde{k}.
\end{equation}

In the present part of this study, the above constitutive matrices, energetically equivalent to the discrete lattice,  
are shown to display non-centrosymmetry and non-local anisotropy, although the local response is isotropic. In particular, the constitutive matrices $\boldsymbol{\mathsf{C}}\left(\overline{k}, \widehat{k}, \widetilde{k}\right)$, $\boldsymbol{\mathsf{M}}\left(\overline{k}, \widehat{k}, \widetilde{k}\right)$, and $\boldsymbol{\mathsf{A}}\left(\overline{k}, \widehat{k}, \widetilde{k}\right)$ are found to belong, respectively, to the symmetry classes $O(2)$ (isotropy), $Z_3$ ($2/3\pi$, without reflection symmetry), and $D_6$ ($\pi/3$, with reflection symmetry).
Furthermore, the positive definiteness of the identified $\mathsf{SGE}$ material is investigated and shown to be possible for specific sets of the bars' stiffness ratios $\widehat{k}/\overline{k}$ and $\widetilde{k}/\overline{k}$.

Finally,  the mechanical response  under simple shear and
uniaxial strain is analytically evaluated for the discrete lattice and for the respective equivalent material. A comparison between the behaviours of the lattice and of its continuous counterpart shows an excellent agreement, a result which provides a validation to the proposed identification procedure and paves the way to practical applications.

\section{From the \lq condensed' to the \lq standard'  $\mathsf{SGE}$ solid: symmetries and positive definiteness}

By means of the energetic matching with a hexagonal lattice (made up of axially deformable bars of length $\ell$ and stiffnesses $\left\{\overline{k}, \widehat{k}, \widetilde{k}\right\}$, Fig. \ref{fig:DisCont}), the strain energy of the equivalent homogeneous second-gradient elastic material ($\mathsf{SGE}$) is given by (in Part I of this study \cite{rizzipt1})

\begin{equation}
\mathcal{U}_{\mathsf{SGE}}\left(\boldsymbol{\mathsf{p}}^{\mathsf{SGE}},\boldsymbol{\mathsf{q}^{*}}\right)  = 
\underbrace{\frac{1}{2}\boldsymbol{\mathsf{p}}^{\mathsf{SGE}^T}\boldsymbol{\mathsf{C}}\left(\overline{k},\widetilde{k},\widehat{k}\right) \boldsymbol{\mathsf{p}}^\mathsf{SGE}}_{\mathcal{U}_{\boldsymbol{\mathsf{C}}}\left(\boldsymbol{\mathsf{p}}^\mathsf{SGE}\right)} + 
\underbrace{\frac{}{}\boldsymbol{\mathsf{p}}^{\mathsf{SGE}^T}\boldsymbol{\mathsf{M}^{*}}\left(\overline{k},\widetilde{k},\widehat{k}\right) \boldsymbol{\mathsf{q}^{*}}}_{\mathcal{U}_{\boldsymbol{\mathsf{M}}^{*}}\left(\boldsymbol{\mathsf{p}}^\mathsf{SGE},\boldsymbol{\mathsf{q}}^{*}\right)} +
\underbrace{\frac{1}{2}\boldsymbol{\mathsf{q}}^{\boldsymbol{*}^T}\boldsymbol{\mathsf{A}^{*}}\left(\overline{k},\widetilde{k},\widehat{k}\right) \boldsymbol{\mathsf{q}^{*}}}_{\mathcal{U}_{\boldsymbol{\mathsf{A}}^{*}}\left(\boldsymbol{\mathsf{q}}^{*}\right)} ,
\label{eq:EneCont2}
\end{equation}
which is a function of the vectors $\boldsymbol{\mathsf{p}}^{\mathsf{SGE}}=\left[\epsilon^{\mathsf{SGE}}_{11},\epsilon^{\mathsf{SGE}}_{22},2\epsilon^{\mathsf{SGE}}_{12}\right]$ and $\boldsymbol{\mathsf{q}^{*}}=\left[\chi^{\mathsf{SGE}}_{111},\chi^{\mathsf{SGE}}_{221},\chi^{\mathsf{SGE}}_{112},\chi^{\mathsf{SGE}}_{222}\right]$, collecting respectively the independent components of the deformation $\bepsilon$ and curvature $\bchi$ (constrained to produce equilibrated stress fields).

The energy $\mathcal{U}_{\mathsf{SGE}}\left(\boldsymbol{\mathsf{p}}^{\mathsf{SGE}},\boldsymbol{\mathsf{q}^{*}}\right)$ also involves the properties of the `condensed' version of the second-gradient elastic material, expressed by the three matrices $\boldsymbol{\mathsf{C}}\left(\overline{k}, \widehat{k}, \widetilde{k}\right)$, $\boldsymbol{\mathsf{M}}^{*}\left(\overline{k}, \widehat{k}, \widetilde{k}\right)$, and $\boldsymbol{\mathsf{A}}^{*}\left(\overline{k}, \widehat{k}, \widetilde{k}\right)$ assume the following expressions
\begin{equation}
\begin{array}{ccc}
	&\boldsymbol{\mathsf{C}} =
\left(
\begin{array}{ccc}
	\mathsf{C}_{12}+2 \mathsf{C}_{33} & \mathsf{C}_{12} & 0 \\
	\mathsf{C}_{12} & \mathsf{C}_{12}+2 \mathsf{C}_{33} & 0 \\
	0 & 0 & \mathsf{C}_{33} \\
\end{array}
\right)
	,
	\hspace{4mm}
	\boldsymbol{\mathsf{M}}^{*} =
	\dfrac{\left(\widehat{k}-\widetilde{k}\right) \left(\widehat{k} \widetilde{k}-2 \overline{k} \left(\widehat{k}+\widetilde{k}\right)\right)\ell}{8 \sqrt{3} I_{[1]} I_{[2]}^2}
	\left(
	\begin{array}{cccc}
		0 & 0 & \mathsf{m}^{*}_{13} & \mathsf{m}^{*}_{14} \\
		0 & 0 & \mathsf{m}^{*}_{13} & \mathsf{m}^{*}_{14} \\
		0 & 0 & 0 &0 \\
	\end{array}
	\right),
	\vspace{4mm}
	\\
	&\boldsymbol{\mathsf{A}}^{*} =\dfrac{\sqrt{3} I_{[3]} \ell^2}{64 I_{[1]}^2 I_{[2]}^4}
	\left(
	\begin{array}{cccc}
		\mathsf{a}^{*}_{11} & \mathsf{a}^{*}_{12} & 0 & 0 \\
		\mathsf{a}^{*}_{12} & \mathsf{a}^{*}_{22} & 0 & 0 \\
		0 & 0 & \mathsf{a}^{*}_{33} & \mathsf{a}^{*}_{34} \\
		0 & 0 & \mathsf{a}^{*}_{34} & \mathsf{a}^{*}_{44} \\
	\end{array}
	\right) ,
\end{array}
\label{eq:mat_ela_coeff}
\end{equation}
where the coefficients involved in the matrices are 
\begin{equation}
\begin{array}{ccc}
\mathsf{C}_{12} = \dfrac{2I_{[1]}I_{[2]} - 9I_{[3]}}{4 \sqrt{3} I_{[2]}},\qquad
\mathsf{C}_{33} =  \dfrac{9I_{[3]}}{4 \sqrt{3} I_{[2]}},
\end{array}
\label{eq:1stcostant}
\end{equation}
\begin{equation}
\begin{array}{ccc}
\mathsf{m}^{*}_{13} = I_{[1]} I_{[2]}-9 I_{[3]},\qquad
\mathsf{m}^{*}_{14} = -3  (I_{[1]} I_{[2]}+3 I_{[3]}),
\end{array}
\label{eq:3stcostant}
\end{equation}
and 
\begin{equation}
	\begin{array}{rcl}
		\mathsf{a}_{12}^{*} &=& \left[ 10 \overline{k}^5 \left( \widehat{k}+\widetilde{k} \right)^3+5 \overline{k}^4 \left( \widehat{k}+\widetilde{k} \right)^2 \left(4 \widehat{k}^2+5 \widehat{k} \widetilde{k}+4 \widetilde{k}^2\right) \right. \\[5mm]
		&& + \overline{k}^3 \left( \widehat{k}+\widetilde{k} \right) \left( 10 \widehat{k}^4-71 \widehat{k}^3 \widetilde{k}-303 \widehat{k}^2 \widetilde{k}^2-71\widehat{k} \widetilde{k}^3+10 \widetilde{k}^4 \right) + 2 \overline{k}^2 \widehat{k} \widetilde{k} \left(6 \widehat{k}^4-9 \widehat{k}^3 \widetilde{k}+641 \widehat{k}^2 \widetilde{k}^2 - 9 \widehat{k} \widetilde{k}^3+6 \widetilde{k}^4\right) \\[4mm]
		&&  \left. - \overline{k} \widehat{k}^2 \widetilde{k}^2 \left( \widehat{k}+\widetilde{k} \right) \left(33 \widehat{k}^2+\widehat{k}\widetilde{k}+33 \widetilde{k}^2\right)-35 \widehat{k}^3 \widetilde{k}^3 \left( \widehat{k}+\widetilde{k} \right)^2 \right], \\[4mm]
		\mathsf{a}_{12}^{*} - \mathsf{a}_{11}^{*} &=& 12 I_{[1]} I_{[2]}\left[ \left(5\overline{k} \left( \widehat{k}+\widetilde{k} \right) -13 \widehat{k} \widetilde{k}\right) \left(\overline{k}^2 \left( \widehat{k}+\widetilde{k} \right) + \overline{k} \left(\widehat{k}^2 + 6 \widehat{k} \widetilde{k} + \widetilde{k}^2 \right) + \widehat{k} \widetilde{k} \left( \widehat{k} + \widetilde{k} \right) \right) \right],\\[4mm]
		\mathsf{a}_{11}^{*} - \mathsf{a}_{22}^{*} &=& 8 I_{[1]} I_{[2]}\left[ \left(13 \widehat{k} \widetilde{k} - 5\overline{k} \left( \widehat{k} + \widetilde{k} \right) \right) \left( \overline{k}^2 \left( \widehat{k} + \widetilde{k} \right) + \overline{k} \left( \widehat{k}^2 + 12 \widehat{k} \widetilde{k} + \widetilde{k}^2 \right) + \widehat{k} \widetilde{k} \left( \widehat{k} + \widetilde{k} \right)\right) \right],\\[4mm]
		\mathsf{a}_{33}^{*} - \mathsf{a}_{22}^{*} &=& \dfrac{2}{3I_{[3]} } \left[  \left( \overline{k}^2 \left( \widehat{k}+\widetilde{k} \right) + \overline{k} \left(\widehat{k}^2 - 6 \widehat{k} \widetilde{k} + \widetilde{k}^2 \right) + \widehat{k} \widetilde{k} \left( \widehat{k} + \widetilde{k} \right) \right)^2 \left(4 \overline{k}^2 \left( \widehat{k} + \widetilde{k} \right)^3 \right.\right.\\[5mm]
		&&\left.\left. - I_{[3]} \left(4 \widehat{k}^2 + 35 \widehat{k} \widetilde{k} + 4\widetilde{k}^2 \right) + \widehat{k}^2 \widetilde{k}^2 \left( \widehat{k} + \widetilde{k} \right) \right) \right],\\[4mm]
		\mathsf{a}_{33}^{*} - \mathsf{a}_{34}^{*} &=& \dfrac{4I_{[1]}I_{[2]}}{3I_{[3]}} \left[ \left( \overline{k}^2 \left( \widehat{k} + \widetilde{k} \right) + \overline{k} \left(\widehat{k}^2 - 6 \widehat{k} \widetilde{k} + \widetilde{k}^2 \right) + \widehat{k} \widetilde{k} \left( \widehat{k} + \widetilde{k} \right) \right) \left(\overline{k}^2 (\widehat{k}+\widetilde{k}) \left(8 \widehat{k}^2 + \widehat{k} \widetilde{k} + 8 \widetilde{k}^2 \right)\right.\right.\\[5mm]
		&&\left.\left. - I_{[3]} \left(8\widehat{k}^2+31 \widehat{k} \widetilde{k} + 8 \widetilde{k}^2 \right) + 2 \widehat{k}^2 \widetilde{k}^2 \left( \widehat{k} + \widetilde{k} \right) \right) \right],\\[4mm]
		\mathsf{a}_{44}^{*} - \mathsf{a}_{33}^{*} &=& \dfrac{8I_{[1]}I_{[2]}}{3I_{[3]}} \left[  \left( \overline{k}^2 \left( \widehat{k} + \widetilde{k} \right) + \overline{k} \left(\widehat{k}^2 + 12 \widehat{k} \widetilde{k} + \widetilde{k}^2 \right) + \widehat{k} \widetilde{k} \left( \widehat{k} + \widetilde{k} \right) \right) \left( \overline{k}^2 \left( \widehat{k} + \widetilde{k} \right) \left(8 \widehat{k}^2 + \widehat{k} \widetilde{k} + 8 \widetilde{k}^2 \right) \right.\right.\\[5mm]
		&& \left.\left. - I_{[3]} \left( 8\widehat{k}^2 + 31 \widehat{k} \widetilde{k} + 8 \widetilde{k}^2 \right) + 2 \widehat{k}^2 \widetilde{k}^2 \left( \widehat{k}+\widetilde{k} \right) \right) \right].
	\end{array}
\end{equation}

On the other hand, the  energy density stored within a \lq standard' $\mathsf{SGE}$ material is
\begin{equation}
\mathcal{U}_{\mathsf{SGE}}\left(\boldsymbol{\mathsf{p}},\boldsymbol{\mathsf{q}}\right) =
\frac{1}{2}\boldsymbol{\mathsf{p}}^{T}\boldsymbol{\mathsf{C}} \boldsymbol{\mathsf{p}} + 
\frac{}{}\boldsymbol{\mathsf{p}}^{T} \boldsymbol{\mathsf{M}} \boldsymbol{\mathsf{q}}+
\frac{1}{2}\boldsymbol{\mathsf{q}}^{T} \boldsymbol{\mathsf{A}} \boldsymbol{\mathsf{q}},
\label{eq:EneCont3}
\end{equation}
which is now constrained to coincide with the energy stored in the \lq condensed' counterpart, Eq.~(\ref{eq:EneCont2}),  when  the same quadratic displacement field (producing an equilibrated stress field) is applied, so that $\boldsymbol{\mathsf{p}}$ and $\boldsymbol{\mathsf{q}}$ are respectively identified with $\boldsymbol{\mathsf{p}}^{\mathsf{SGE}}$ and $\boldsymbol{\mathsf{q}}^{\mathsf{SGE}}$. In other words, given the matrices $\boldsymbol{\mathsf{M}^{*}} $ and $\boldsymbol{\mathsf{A}^{*}}$, Eq. (\ref{eq:mat_ela_coeff}), the matrices $\boldsymbol{\mathsf{M}}$ and $\boldsymbol{\mathsf{A}}$ are to be determined, which satisfy 
the following relations
\begin{equation} 
\boldsymbol{\mathsf{M}^{*}} = \boldsymbol{\mathsf{M}}\boldsymbol{\mathsf{T}}^{\mathsf{SGE}},
\qquad
\boldsymbol{\mathsf{A}^{*}}=\boldsymbol{\mathsf{T}}^{\mathsf{SGE}^T}\boldsymbol{\mathsf{A}}\boldsymbol{\mathsf{T}}^{\mathsf{SGE}},\qquad
\boldsymbol{\mathsf{q}}^{\mathsf{SGE}} = \boldsymbol{\mathsf{T}}^{\mathsf{SGE}} \boldsymbol{\mathsf{q}}^{*},
\label{eq:toTheStar}
\end{equation}
where the transformation matrix $\boldsymbol{\mathsf{T}}^{\mathsf{SGE}}$, providing the equilibrium conditions in the second-gradient elastic solid, is introduced
\begin{equation}
\boldsymbol{\mathsf{T}}^{\mathsf{SGE}} =
\left(
\begin{array}{cccc}
1 & 0 & 0 & 0 \\
0 & 1 & 0 & 0 \\
0 & 0 & 1 & 0 \\
0 & 0 & 0 & 1 \\[3mm]
0 & 0 & -\dfrac{2 \mathsf{C}_{33}}{\mathsf{C}_{12}+\mathsf{C}_{33}} & \dfrac{2 \mathsf{C}_{12}}{\mathsf{C}_{12}+\mathsf{C}_{33}}-4 \\[3mm]
\dfrac{2 \mathsf{C}_{12}}{\mathsf{C}_{12}+\mathsf{C}_{33}}-4 & -\dfrac{2 \mathsf{C}_{33}}{\mathsf{C}_{12}+\mathsf{C}_{33}} & 0 & 0 \\
\end{array}
\right).
\label{eq:q_star}
\end{equation}

From Eqs.~(\ref{eq:toTheStar}), the matrices $\boldsymbol{\mathsf{M}}$ and  $\boldsymbol{\mathsf{A}}$  can be obtained from the identified matrices $\boldsymbol{\mathsf{M}^{*}}$ and $\boldsymbol{\mathsf{A}^{*}}$, and the vector $\boldsymbol{\mathsf{q}}^{*}$ can be related to the vector $\boldsymbol{\mathsf{q}}^{\mathsf{SGE}}$ through the following linear relations
\begin{equation}
\boldsymbol{\mathsf{M}} = \boldsymbol{\mathsf{M}^{*}}\boldsymbol{\mathsf{Q}} + \Delta \boldsymbol{\mathsf{M}},
\quad
\boldsymbol{\mathsf{A}}=\boldsymbol{\mathsf{Q}}^{T}\boldsymbol{\mathsf{A}^{*}}\boldsymbol{\mathsf{Q}} + \Delta \boldsymbol{\mathsf{A}}
\quad \mbox{and} \quad
\boldsymbol{\mathsf{q}^{*}} = \boldsymbol{\mathsf{Q}} \boldsymbol{\mathsf{q}}^{\mathsf{SGE}},
\label{eq:andBack}
\end{equation}
where $\Delta \boldsymbol{\mathsf{M}}$ and $\Delta \boldsymbol{\mathsf{A}}$ are additional constitutive matrices (the latter symmetric) yet to be identified, while $\boldsymbol{\mathsf{Q}}$ is a rectangular $4 \times 6$ transformation matrix, with the properties
\begin{equation}
\boldsymbol{\mathsf{Q}}\,\boldsymbol{\mathsf{T}}^{\mathsf{SGE}}=\bI,
\qquad
\Delta\boldsymbol{\mathsf{M}}\,\boldsymbol{\mathsf{T}}^{\mathsf{SGE}}=\b0,
\qquad
\Delta\boldsymbol{\mathsf{A}}\,\boldsymbol{\mathsf{T}}^{\mathsf{SGE}}=\b0,
\label{eq:constrain_Q_DM_DA}
\end{equation}
(being $\bI$ the identity matrix of rank 4) so that the energy density of the \lq standard' $\mathsf{SGE}$ material is equal to that of the  \lq condensed' form. It is noted that the constraint provided by Eqs.~(\ref{eq:constrain_Q_DM_DA}) does not completely define the matrices $\boldsymbol{\mathsf{Q}}$, $\Delta\boldsymbol{\mathsf{M}}$ and $\Delta\boldsymbol{\mathsf{A}}$.
Indeed,  eight over the twenty-four coefficients of the transformation matrix $\boldsymbol{\mathsf{Q}}$, six over the eighteen coefficients of $\Delta\boldsymbol{\mathsf{M}}$ and three over the twenty-one independent coefficient of $\Delta\boldsymbol{\mathsf{A}}$ remain still undetermined after imposing Eqs.~(\ref{eq:constrain_Q_DM_DA}).
These matrices can be expressed as
\begin{equation}
\begin{array}{ccc}
&\boldsymbol{\mathsf{Q}} =
\resizebox{0.85\textwidth}{!}{$
-\mathsf{Z}
\left(
\begin{array}{cccccc}
\mathsf{Q}_{16} & 0 & 0 & \mathsf{Q}_{15} & 0 & 0 \\[3mm]
\mathsf{Q}_{26} & 0 & 0 & \mathsf{Q}_{25} & 0 & 0 \\[3mm]
\mathsf{Q}_{36} & 0 & 0 & \mathsf{Q}_{35} & 0 & 0 \\[3mm]
\mathsf{Q}_{46} & 0 & 0 & \mathsf{Q}_{45} & 0 & 0 \\
\end{array}
\right) -
\mathsf{Y}
\left(
\begin{array}{cccccc}
0 & \mathsf{Q}_{16} & \mathsf{Q}_{15} & 0 & 0 & 0 \\[3mm]
0 & \mathsf{Q}_{26} & \mathsf{Q}_{25} & 0 & 0 & 0 \\[3mm]
0 & \mathsf{Q}_{36} & \mathsf{Q}_{35} & 0 & 0 & 0 \\[3mm]
0 & \mathsf{Q}_{46} & \mathsf{Q}_{45} & 0 & 0 & 0 \\
\end{array}
\right) +
\left(
\begin{array}{cccccc}
1 & 0 & 0 & 0 & \mathsf{Q}_{15} & \mathsf{Q}_{16} \\[3mm]
0 & 1 & 0 & 0 & \mathsf{Q}_{25} & \mathsf{Q}_{26} \\[3mm]
0 & 0 & 1 & 0 & \mathsf{Q}_{35} & \mathsf{Q}_{36} \\[3mm]
0 & 0 & 0 & 1 & \mathsf{Q}_{45} & \mathsf{Q}_{46} \\
\end{array}
\right)
$},
\vspace{3mm}\\
&\Delta\boldsymbol{\mathsf{M}} = 
\left(
\begin{array}{cccccc}
-\Delta\mathsf{M}_{16} \mathsf{Z} & -\Delta\mathsf{M}_{16} \mathsf{Y} & -\Delta\mathsf{M}_{15} \mathsf{Y} & -\Delta\mathsf{M}_{15} \mathsf{Z} & \Delta\mathsf{M}_{15} & \Delta\mathsf{M}_{16} \\
-\Delta\mathsf{M}_{26} \mathsf{Z} & -\Delta\mathsf{M}_{26} \mathsf{Y} & -\Delta\mathsf{M}_{25} \mathsf{Y} & -\Delta\mathsf{M}_{25} \mathsf{Z} & \Delta\mathsf{M}_{25} & \Delta\mathsf{M}_{26} \\
-\Delta\mathsf{M}_{36} \mathsf{Z} & -\Delta\mathsf{M}_{36} \mathsf{Y} & -\Delta\mathsf{M}_{35} \mathsf{Y} & -\Delta\mathsf{M}_{35} \mathsf{Z} & \Delta\mathsf{M}_{35} & \Delta\mathsf{M}_{36} \\
\end{array}
\right),
\vspace{3mm}\\
&\Delta\boldsymbol{\mathsf{A}} = 
\left(
\begin{array}{cccccc}
\Delta\mathsf{A}_{11} & \dfrac{\Delta\mathsf{A}_{11} \mathsf{Y}}{\mathsf{Z}} & -\Delta\mathsf{A}_{15} \mathsf{Y} & -\Delta\mathsf{A}_{15} \mathsf{Z} & \Delta\mathsf{A}_{15} & -\dfrac{\Delta\mathsf{A}_{11}}{\mathsf{Z}} \\
\dfrac{\Delta\mathsf{A}_{11} \mathsf{Y}}{\mathsf{Z}} & \dfrac{\Delta\mathsf{A}_{11} \mathsf{Y}^2}{\mathsf{Z}^2} & -\dfrac{\Delta\mathsf{A}_{15} \mathsf{Y}^2}{\mathsf{Z}} & -\Delta\mathsf{A}_{15} \mathsf{Y} &
\dfrac{\Delta\mathsf{A}_{15} \mathsf{Y}}{\mathsf{Z}} & -\dfrac{\Delta\mathsf{A}_{11} \mathsf{Y}}{\mathsf{Z}^2} \\
-\Delta\mathsf{A}_{15} \mathsf{Y} & -\dfrac{\Delta\mathsf{A}_{15} \mathsf{Y}^2}{\mathsf{Z}} & -\Delta\mathsf{A}_{35} \mathsf{Y} & -\Delta\mathsf{A}_{35} \mathsf{Z} & \Delta\mathsf{A}_{35} & \dfrac{\Delta\mathsf{A}_{15}
	\mathsf{Y}}{\mathsf{Z}} \\
-\Delta\mathsf{A}_{15} \mathsf{Z} & -\Delta\mathsf{A}_{15} \mathsf{Y} & -\Delta\mathsf{A}_{35} \mathsf{Z} & -\dfrac{\Delta\mathsf{A}_{35} \mathsf{Z}^2}{\mathsf{Y}} & \dfrac{\Delta\mathsf{A}_{35} \mathsf{Z}}{\mathsf{Y}} &
\Delta\mathsf{A}_{15} \\
\Delta\mathsf{A}_{15} & \dfrac{\Delta\mathsf{A}_{15} \mathsf{Y}}{\mathsf{Z}} & \Delta\mathsf{A}_{35} & \dfrac{\Delta\mathsf{A}_{35} \mathsf{Z}}{\mathsf{Y}} & -\dfrac{\Delta\mathsf{A}_{35}}{\mathsf{Y}} &
-\dfrac{\Delta\mathsf{A}_{15}}{\mathsf{Z}} \\
-\dfrac{\Delta\mathsf{A}_{11}}{\mathsf{Z}} & -\dfrac{\Delta\mathsf{A}_{11} \mathsf{Y}}{\mathsf{Z}^2} & \dfrac{\Delta\mathsf{A}_{15} \mathsf{Y}}{\mathsf{Z}} & \Delta\mathsf{A}_{15} &
-\dfrac{\Delta\mathsf{A}_{15}}{\mathsf{Z}} & \dfrac{\Delta\mathsf{A}_{11}}{\mathsf{Z}^2} \\
\end{array}
\right),
\end{array}
\label{eq:Q_DM_DA}
\end{equation}
where $\mathsf{Y}=-2 \mathsf{C}_{33}/(\mathsf{C}_{12}+\mathsf{C}_{33})$ and $\mathsf{Z}=-2 (\mathsf{C}_{12}+2 \mathsf{C}_{33})/(\mathsf{C}_{12}+\mathsf{C}_{33})$.

In the following of this section it is first shown that the three matrices $\boldsymbol{\mathsf{C}}\left(\overline{k}, \widehat{k}, \widetilde{k}\right)$, $\boldsymbol{\mathsf{M}}^{*}\left(\overline{k}, \widehat{k}, \widetilde{k}\right)$, and $\boldsymbol{\mathsf{A}}^{*}\left(\overline{k}, \widehat{k}, \widetilde{k}\right)$ 
(identified in Part I of this study) 
respectively belong to the symmetry  classes $O(2)$ (isotropy), $Z_3$ ($2/3\pi$, without reflection symmetry), and $D_6$ ($\pi/3$, with reflection symmetry).
These symmetry properties are then used to further constraint the undetermined coefficients of the matrices $\boldsymbol{\mathsf{Q}}$, $\Delta\boldsymbol{\mathsf{M}}$, and  $\Delta\boldsymbol{\mathsf{A}}$, a procedure which will leave undetermined 1 component of 
$\boldsymbol{\mathsf{Q}}$, 2 components of $\Delta\boldsymbol{\mathsf{M}}$, and  1 component of $\Delta\boldsymbol{\mathsf{A}}$. The remaining 4 components will be eventually 
determined with a \lq practical' rule, so that the second-gradient elastic solid equivalent to the lattice will be defined. 
With this definition, the domain of positive definiteness of both the equivalent material in the standard and condensed forms will be the same, at varying of the two independent stiffness ratios of the lattice. 

\subsection{Anisotropy characterization}
\label{sec:Aniso}

Considering an orthogonal matrix $\bD$ which realizes a generic rotation or reflection ($\bD \bD^T = \bD^T\bD=\bI$), the  components of strain and curvature in the rotated or reflected reference system $\{\epsilon_{ij}^{\#},\chi_{ijk}^{\#}\}$  are related  to those in the untrasformed system $\{\epsilon_{ij},\chi_{ijk}\}$ by \cite{AUFFRAY20101698, auffray2015complete}
\begin{equation}
	\begin{array}{ccc}
	\epsilon^{\#}_{ij}=\epsilon_{kl}D_{ik}D_{jl}, & \chi^{\#}_{ijk}= \chi_{lmn} D_{il}D_{jm}D_{kn}.
	\label{eq:trasfor1}
	\end{array}
\end{equation}
$\bD$ will be identified either with a reflection $\bS$ and with a rotation  $\bR\left(\theta\right)$. These, restricted to a two-dimensional space, can be expressed as
\begin{equation}
\bS=
\left(
\begin{array}{ccc}
1 & 0  \\
0 & -1 \\
\end{array}
\right),
\qquad
\bR =
\left(
\begin{array}{ccc}
c(\theta) & -s(\theta) \\
s(\theta) & c(\theta)
\end{array}
\right),
\label{eq:mat_rota}
\end{equation}
with  $s(\theta)=\sin\theta$ and $c(\theta)=\cos\theta$.
When the strain and curvature tensors are represented as vectors in the Voigt notation, Eqs. (\ref{eq:trasfor1}) have to be rewritten in the following form 
\begin{equation}
\begin{array}{ccc}
\boldsymbol{\mathsf{p}}^{\mathsf{SGE}^\#} =  \boldsymbol{\mathsf{D}}_{[\boldsymbol{\mathsf{p}}]} \boldsymbol{\mathsf{p}}^{\mathsf{SGE}},&
\boldsymbol{\mathsf{q}}^{\mathsf{SGE}^\#} =  \boldsymbol{\mathsf{D}}_{[\boldsymbol{\mathsf{q}}]} \boldsymbol{\mathsf{q}}^{\mathsf{SGE}},
\end{array}
\label{eq:chi_rota}
\end{equation}
where the pairs of matrices $\boldsymbol{\mathsf{D}_{[\boldsymbol{\mathsf{p}}]}}$ and $\boldsymbol{\mathsf{D}_{[\boldsymbol{\mathsf{q}}]}}$ will be taken in the following to coincide respectively with either the pair 
$\boldsymbol{\mathsf{S}}_{[\boldsymbol{\mathsf{p}}]}$ and $\boldsymbol{\mathsf{S}}_{[\boldsymbol{\mathsf{q}}]}$ or 
$\boldsymbol{\mathsf{R}}_{[\boldsymbol{\mathsf{p}}]}$ 
and
$\boldsymbol{\mathsf{R}}_{[\boldsymbol{\mathsf{q}}]}$, defined as 
\begin{equation}
\begin{array}{c}
\boldsymbol{\mathsf{S}}_{[\boldsymbol{\mathsf{p}}]}=
\left(
\begin{array}{ccc}
1 & 0 & 0 \\
0 & 1 & 0 \\
0 & 0 & -1 
\end{array}
\right),
\quad
\boldsymbol{\mathsf{S}}_{[\boldsymbol{\mathsf{q}}]}=
\resizebox{0.2\textwidth}{!}{$
	\left(
	\begin{array}{cccccc}
	1 & 0 & 0 & 0 & 0 & 0 \\
	0 & 1 & 0 & 0 & 0 & 0 \\
	0 & 0 & -1 & 0 & 0 & 0 \\
	0 & 0 & 0 & -1 & 0 & 0 \\
	0 & 0 & 0 & 0 & -1 & 0 \\
	0 & 0 & 0 & 0 & 0 & 1 
	\end{array}
	\right)
	$},
\quad
\boldsymbol{\mathsf{R}}_{[\boldsymbol{\mathsf{p}}]}(\theta)=
\left(
\begin{array}{ccc}
c ^2(\theta ) & s ^2(\theta ) & -c (\theta ) s (\theta ) \\
s ^2(\theta ) & c ^2(\theta ) & c (\theta ) s (\theta ) \\
s (2 \theta ) & -2 c (\theta ) s (\theta ) & c (2 \theta )
\end{array}
\right),
\vspace{5mm}
\\
\boldsymbol{\mathsf{R}}_{[\boldsymbol{\mathsf{q}}]}(\theta)=
\left(
\begin{array}{cccccc}
c ^3(\theta ) & s ^2(\theta ) c (\theta ) & - s (\theta ) c ^2(\theta ) & -s ^3(\theta ) & - s (\theta ) c ^2(\theta) & s ^2(\theta ) c (\theta ) \\
s ^2(\theta ) c (\theta ) & c ^3(\theta ) & -s ^3(\theta ) & - s (\theta ) c ^2(\theta ) & s (\theta ) c ^2(\theta ) &	- s ^2(\theta ) c (\theta ) \\
s (\theta ) c ^2(\theta ) & s ^3(\theta ) & c ^3(\theta ) & s ^2(\theta ) c (\theta ) & -s ^2(\theta ) c (\theta ) & - s (\theta ) c ^2(\theta ) \\
s ^3(\theta ) & s (\theta ) c ^2(\theta ) & s ^2(\theta ) c (\theta ) & c ^3(\theta ) & s ^2(\theta ) c (\theta ) & s (\theta )	c ^2(\theta ) \\
2 s (\theta ) c ^2(\theta ) & -2 s (\theta ) c ^2(\theta ) & -2 s ^2(\theta ) c (\theta ) & s (\theta ) s (2 \theta ) & (c (\theta )+c (3 \theta ))/2 & (s (\theta )-s (3 \theta ))/2 \\
s (\theta ) s (2 \theta ) & -2 s ^2(\theta ) c (\theta ) & 2 s (\theta ) c ^2(\theta ) & -2 s (\theta ) c ^2(\theta ) & s (\theta) c (2 \theta ) & (c (\theta )+c (3 \theta ))/2.
\end{array}
\right)
\end{array}
\label{eq:mat_rot_non_conde}
\end{equation}
Considering the above premise, the symmetry classes of the constitutive matrices can be investigated. While the symmetry analysis for the first-order response can be immediately performed, the same analysis for the higher-order response requires the definition of the transformation matrices in condensed form related to the presence of the curvature vector $\boldsymbol{\mathsf{q}^{*}}$.

\paragraph{Isotropy of the first-order equivalent material $\boldsymbol{\mathsf{C}}$.}
The first-order response is symmetric with respect to the rotation $\theta$ when 
\begin{equation}
	 \boldsymbol{\mathsf{R}}_{[\boldsymbol{\mathsf{p}}]}^{T}(\theta) \boldsymbol{\mathsf{C}} \boldsymbol{\mathsf{R}}_{[\boldsymbol{\mathsf{p}}]} (\theta)  =  \boldsymbol{\mathsf{C}}, 
	\label{eq:symmetries_C}
\end{equation}
and mirror symmetric with respect to the reflection $\boldsymbol{\mathsf{S}_{[\boldsymbol{\mathsf{p}}]}}$ when
\begin{equation}
	 	\boldsymbol{\mathsf{S}}_{[\boldsymbol{\mathsf{p}}]}^{T} \boldsymbol{\mathsf{C}} \boldsymbol{\mathsf{S}}_{[\boldsymbol{\mathsf{p}}]}  =  \boldsymbol{\mathsf{C}}.
\label{eq:symmetries_C_2}
\end{equation}

Holding true these conditions for every $\theta$, the local response for the equivalent material expressed by $\boldsymbol{\mathsf{C}}$, Eq.~(\ref{eq:mat_ela_coeff})$_1$, is isotropic and centrosymmetric.

\subsubsection{Anisotropy characterization of $\boldsymbol{\mathsf{M}}^{*}$ and $\boldsymbol{\mathsf{A}}^{*}$}

In a reference system denoted with symbol $\#$ the following relations hold
\begin{equation}
\begin{array}{ccc}
\boldsymbol{\mathsf{q}}^{\mathsf{SGE}^{\#}} = \boldsymbol{\mathsf{T}}^{\mathsf{SGE}^\#} \boldsymbol{\mathsf{q}}^{*^\#}, &  ~ & \boldsymbol{\mathsf{q}}^{*^\#} = \boldsymbol{\mathsf{Q}}^{\#} \boldsymbol{\mathsf{q}}^{\mathsf{SGE}^{\#}} ,
\end{array}
\label{eq:chi_contrac_rota}
\end{equation}
so that the transformed Eq. (\ref{eq:constrain_Q_DM_DA}) can now be written as 
\begin{equation}
\boldsymbol{\mathsf{Q}}^{\#}\boldsymbol{\mathsf{T}}^{\mathsf{SGE}^\#}=\bI, 
\label{eq:constrain_Q_Qash}
\end{equation}
while the condensed vector  $\boldsymbol{\mathsf{q}}^{*}$ transforms according to the rule
\begin{equation}
\boldsymbol{\mathsf{q}}^{*^\#} =  \boldsymbol{\mathsf{D}}_{[\boldsymbol{\mathsf{q}}^{*}]}^{*} \boldsymbol{\mathsf{q}}^{*}, 
\label{eq:trasfor3}
\end{equation}
where $\boldsymbol{\mathsf{D}}_{[\boldsymbol{\mathsf{q}}^{*}]}^{*} $ is the condensed version of $\boldsymbol{\mathsf{D}}_{[\boldsymbol{\mathsf{q}}]}$ and can be found as follows, with the premise that the above equations simplify because the first-order material is isotropic, so that 
\begin{equation}
\begin{array}{ll}
\boldsymbol{\mathsf{T}}^{\mathsf{SGE}^\#} = \boldsymbol{\mathsf{T}}^{\mathsf{SGE}}, & \boldsymbol{\mathsf{Q}} = \boldsymbol{\mathsf{Q}}^{\#} .
\end{array}
\label{eq:sempli_iso}
\end{equation}

Substituting Eq.~(\ref{eq:toTheStar})$_3$ into Eq.~(\ref{eq:chi_rota}) and pre-multiplying the obtained equation by the transformation matrix $\boldsymbol{\mathsf{Q}}$ leads to
\begin{equation}
\boldsymbol{\mathsf{Q}} \boldsymbol{\mathsf{q}}^{\mathsf{SGE}^\#} = \boldsymbol{\mathsf{Q}} \boldsymbol{\mathsf{D}}_{[\boldsymbol{\mathsf{q}}]} \boldsymbol{\mathsf{T}}^{\mathsf{SGE}} \boldsymbol{\mathsf{q}}^{*},
\label{eq:chi_rota_vs_contra}
\end{equation}
so that the left side of Eq. (\ref{eq:chi_rota_vs_contra}) is the transformed condensed curvature vector $\boldsymbol{\mathsf{q}}^{*^\#}$, Eq. (\ref{eq:chi_contrac_rota})$_2$. The latter equation, compared with Eq.~(\ref{eq:trasfor3}), yields
\begin{equation}
\boldsymbol{\mathsf{D}}^{*}_{[\boldsymbol{\mathsf{q}}^{*}]} = \boldsymbol{\mathsf{Q}} \boldsymbol{\mathsf{D}}_{[\boldsymbol{\mathsf{q}}]} \boldsymbol{\mathsf{T}}^{\mathsf{SGE}},
\label{eq:chi_rotacontra_contra}
\end{equation}
providing the relation between $\boldsymbol{\mathsf{D}}_{[\boldsymbol{\mathsf{q}}^{*}]}^{*} $  and $\boldsymbol{\mathsf{D}}_{[\boldsymbol{\mathsf{q}}]}$.

A replacement of $\boldsymbol{\mathsf{D}}_{[\boldsymbol{\mathsf{q}}]}$ by the rotation $\boldsymbol{\mathsf{R}}_{[\boldsymbol{\mathsf{q}}]}$ or by the reflection $\boldsymbol{\mathsf{S}}_{[\boldsymbol{\mathsf{q}}]}$  in the law (\ref{eq:chi_rotacontra_contra}) 
provides the condensed rotation $\boldsymbol{\mathsf{R}}_{[\boldsymbol{\mathsf{q}}^{*}]}^{*}$ and reflection $\boldsymbol{\mathsf{S}}_{[\boldsymbol{\mathsf{q}}^{*}]}^{*}$ as
\begin{equation}
\boldsymbol{\mathsf{R}}_{[\boldsymbol{\mathsf{q}}^{*}]}^{*}(\theta)=
\resizebox{0.88\textwidth}{!}{$
\left(
\begin{array}{cccc}
\dfrac{c(\theta ) ((3 \mathsf{C}_{12}+5 \mathsf{C}_{33}) c(2 \theta )-\mathsf{C}_{12}-3 \mathsf{C}_{33})}{2 (\mathsf{C}_{12}+\mathsf{C}_{33})} &
\dfrac{(\mathsf{C}_{12}-\mathsf{C}_{33}) s ^2(\theta ) c(\theta )}{\mathsf{C}_{12}+\mathsf{C}_{33}} & \dfrac{(\mathsf{C}_{33}-\mathsf{C}_{12}) s (\theta ) c^2(\theta
	)}{\mathsf{C}_{12}+\mathsf{C}_{33}} & \dfrac{s (\theta ) ((3 \mathsf{C}_{12}+5 \mathsf{C}_{33}) c(2 \theta )+\mathsf{C}_{12}+3 \mathsf{C}_{33})}{2 (\mathsf{C}_{12}+\mathsf{C}_{33})}
\\
\dfrac{(3 \mathsf{C}_{12}+5 \mathsf{C}_{33}) s ^2(\theta ) c(\theta )}{\mathsf{C}_{12}+\mathsf{C}_{33}} & \dfrac{c(\theta ) ((\mathsf{C}_{12}-\mathsf{C}_{33}) c(2
	\theta )+\mathsf{C}_{12}+3 \mathsf{C}_{33})}{2 (\mathsf{C}_{12}+\mathsf{C}_{33})} & \dfrac{s (\theta ) ((\mathsf{C}_{12}-\mathsf{C}_{33}) c(2 \theta )-\mathsf{C}_{12}-3
	\mathsf{C}_{33})}{2 (\mathsf{C}_{12}+\mathsf{C}_{33})} & -\dfrac{(3 \mathsf{C}_{12}+5 \mathsf{C}_{33}) s (\theta ) c^2(\theta )}{\mathsf{C}_{12}+\mathsf{C}_{33}} \\
\dfrac{(3 \mathsf{C}_{12}+5 \mathsf{C}_{33}) s (\theta ) c^2(\theta )}{\mathsf{C}_{12}+\mathsf{C}_{33}} & \dfrac{s (\theta ) ((\mathsf{C}_{33}-\mathsf{C}_{12}) c(2
	\theta )+\mathsf{C}_{12}+3 \mathsf{C}_{33})}{2 (\mathsf{C}_{12}+\mathsf{C}_{33})} & \dfrac{c(\theta ) ((\mathsf{C}_{12}-\mathsf{C}_{33}) c(2 \theta )+\mathsf{C}_{12}+3
	\mathsf{C}_{33})}{2 (\mathsf{C}_{12}+\mathsf{C}_{33})} & \dfrac{(3 \mathsf{C}_{12}+5 \mathsf{C}_{33}) s ^2(\theta ) c(\theta )}{\mathsf{C}_{12}+\mathsf{C}_{33}} \\
-\dfrac{s (\theta ) ((3 \mathsf{C}_{12}+5 \mathsf{C}_{33}) c(2 \theta )+\mathsf{C}_{12}+3 \mathsf{C}_{33})}{2 (\mathsf{C}_{12}+\mathsf{C}_{33})} &
\dfrac{(\mathsf{C}_{12}-\mathsf{C}_{33}) s (\theta ) c^2(\theta )}{\mathsf{C}_{12}+\mathsf{C}_{33}} & \dfrac{(\mathsf{C}_{12}-\mathsf{C}_{33}) s ^2(\theta ) c(\theta
	)}{\mathsf{C}_{12}+\mathsf{C}_{33}} & \dfrac{c(\theta ) ((3 \mathsf{C}_{12}+5 \mathsf{C}_{33}) c(2 \theta )-\mathsf{C}_{12}-3 \mathsf{C}_{33})}{2 (\mathsf{C}_{12}+\mathsf{C}_{33})}
\\
\end{array}
\right)
$},
\end{equation}

\begin{equation}
\boldsymbol{\mathsf{S}}_{[\boldsymbol{\mathsf{q}}^{*}]}^{*}=
\left(
\begin{array}{cccc}
1 & 0 & 0 & 0 \\
0 & 1 & 0 & 0 \\
0 & 0 & -1 & 0 \\
0 & 0 & 0 & -1 \\
\end{array}
\right).
\end{equation}

Note that both $\boldsymbol{\mathsf{R}}_{[\boldsymbol{\mathsf{q}}^{*}]}^{*}(\theta)$ and $\boldsymbol{\mathsf{S}}_{[\boldsymbol{\mathsf{q}}^{*}]}^{*}$ do not depend on the coefficients of $\boldsymbol{\mathsf{Q}}$, a property which follows from the isotropy of the first-order effective material $\boldsymbol{\mathsf{C}}$.

\paragraph{Anisotropy characterization of $\boldsymbol{\mathsf{M}}^{*}$.}
The matrix $\boldsymbol{\mathsf{M}}^{*}$ is symmetric with respect to the rotation $\theta$ when 
\begin{equation}
\boldsymbol{\mathsf{R}_{[\boldsymbol{\mathsf{p}}]}}^{T}(\theta) \boldsymbol{\mathsf{M}}^{*} \boldsymbol{\mathsf{R}_{[\boldsymbol{\mathsf{q}}^{*}]}}^{*}(\theta)  =  \boldsymbol{\mathsf{M}}^{*}, 
\label{eq:symmetries_M}
\end{equation}
and mirror symmetric with respect to the reflection $\boldsymbol{\mathsf{S}_{[\boldsymbol{\mathsf{p}}]}}$ when
\begin{equation}
\boldsymbol{\mathsf{S}_{[\boldsymbol{\mathsf{p}}]}}^{T} \boldsymbol{\mathsf{M}}^{*} \boldsymbol{\mathsf{S}_{[\boldsymbol{\mathsf{q}}^{*}]}}^{*}  =  \boldsymbol{\mathsf{M}}^{*}.
\label{eq:symmetries_M_2}
\end{equation}

For the  matrix $\boldsymbol{\mathsf{M}}^{*}$ defined by Eq.~(\ref{eq:mat_ela_coeff})$_2$, the condition (\ref{eq:symmetries_M}) is verified for $\theta = 2j\pi/3$ ($j\in\mathbb{Z}$), so that the material belongs to $Z_3$ class of symmetry. Moreover, the material is non-centrosymmetric being Eq.~(\ref{eq:symmetries_M_2}) in general not verified. 
The sole exception is verifed when
\begin{equation}
\left(\widehat{k}-\widetilde{k}\right) \left(\widehat{k} \widetilde{k}-2 \overline{k} \left(\widehat{k}+\widetilde{k}\right)\right)\ell=0,
\label{eq:M0}
\end{equation}
corresponding to a centrosymmetric response, $\boldsymbol{\mathsf{M}}^{*}=0$. Such a condition is attained whenever the bars stiffness satisfy one of the two conditions
\begin{equation}
\widehat{k}=\widetilde{k} \quad \mbox{ or } \quad \overline{k}=\dfrac{\widehat{k} \widetilde{k}}{2 \left(\widehat{k} + \widetilde{k}\right)}.
\label{eq:constrain_CS}
\end{equation}

\paragraph{Anisotropy characterization of $\boldsymbol{\mathsf{A}}^{*}$.}
The matrix $\boldsymbol{\mathsf{A}}^{*}$ is symmetric with respect to the rotation $\theta$ when 
\begin{equation}
\boldsymbol{\mathsf{R}}_{[\boldsymbol{\mathsf{q}}^*]}^{*^T}(\theta) \boldsymbol{\mathsf{A}}^{*} \boldsymbol{\mathsf{R}}_{[\boldsymbol{\mathsf{q}}^{*}]}^{*}(\theta)  =  \boldsymbol{\mathsf{A}}^{*}, 
\label{eq:symmetries_A}
\end{equation}
and mirror symmetric with respect to the reflection $\boldsymbol{\mathsf{S}_{[\boldsymbol{\mathsf{p}}]}}$ when
\begin{equation}
\boldsymbol{\mathsf{S}_{[\boldsymbol{\mathsf{q}}^*]}}^{*^T} \boldsymbol{\mathsf{A}}^{*} \boldsymbol{\mathsf{S}_{[\boldsymbol{\mathsf{q}}^{*}]}}^{*}  =  \boldsymbol{\mathsf{A}}^{*}.
\label{eq:symmetries_A_2}
\end{equation}

For the matrix $\boldsymbol{\mathsf{A}}^{*}$ defined by Eq.~(\ref{eq:mat_ela_coeff})$_3$,  condition (\ref{eq:symmetries_A}) is verified for $\theta = j\pi/3$ ($j\in\mathbb{Z}$), while  condition (\ref{eq:symmetries_A_2}) always is, so that the material belongs to $D_6$ class of symmetry.
To annihilate the anisotropic component of the matrix $\boldsymbol{\mathsf{A}}^{*}$, the following condition for the three bars stiffness has to be realized
\begin{equation}
\overline{k} = \dfrac{\widehat{k} \widetilde{k} \left(4 \widehat{k}^2+35 \widehat{k} \widetilde{k}+4 \widetilde{k}^2\right) \pm 3 \sqrt{3} \sqrt{\widehat{k}^3 \widetilde{k}^3 \left(8 \widehat{k}^2+43 \widehat{k} \widetilde{k}+8 \widetilde{k}^2\right)}}{8 \left(\widehat{k}+\widetilde{k}\right)^3}.
\label{eq:A0}
\end{equation}

\subsubsection{Isotropic and centrosymmetric second-gradient equivalent material}
The only possibility of obtaining a second-gradient elastic material characterized by an isotropic response (also implying centrosymmetry) occurs when both Eqs. (\ref{eq:M0}) and (\ref{eq:A0}) are satisfied, which leads to the following constraint on $\overline{k}, \widehat{k}$, and $\widetilde{k}$
\begin{equation}
\widehat{k}=\widetilde{k} \quad \mbox{ and } \quad \overline{k}=\dfrac{43 \pm 3\sqrt{177}}{64}\widetilde{k}.
\label{eq:Isotropy}
\end{equation}

The symmetries of the \lq condensed' material are now completely characterized. As shown in Sect.~\ref{Sec:relaxed}, requiring such symmetry properties to hold for the \lq standard' material provides a further constraint for the remaining free parameters in the matrices $\boldsymbol{\mathsf{Q}}$, $\Delta\boldsymbol{\mathsf{M}}$, and $\Delta\boldsymbol{\mathsf{A}}$, Eq.~(\ref{eq:Q_DM_DA}), with the exception of $\mathsf{Q}_{35}$, $\Delta\mathsf{M}_{15}$, $\Delta\mathsf{M}_{16}$, and $\Delta\mathsf{A}_{11}$, which remain undetermined. 

\subsection{The directional properties of the elastic energy for the $\mathsf{SGE}$ material}

Inspired by \cite{lekhnitskii1964theory}, a further insight on the anisotropy of the effective $\mathsf{SGE}$ material (still in the \lq condensed' form) can be provided by analyzing the effect of a rotation of the applied curvature field on the stored energy density $\mathcal{U}_{\mathsf{SGE}}$.
The curvature part $\mathcal{U}_{\boldsymbol{\mathsf{A}}^{*}}$ of the elastic energy density $\mathcal{U}_{\mathsf{SGE}}$, Eq.~(\ref{eq:EneCont2}), generated by the rotation matrix $\boldsymbol{\mathsf{R}}_{[\boldsymbol{\mathsf{q}}^*]}^{*^T}(\theta)$ applied to a curvature $\boldsymbol{\mathsf{q}}^{*}$, is
\begin{equation}
\mathcal{U}_{\boldsymbol{\mathsf{A}}^{*}}\left(\boldsymbol{\mathsf{q}}^{*},\theta\right)=\frac{1}{2} \boldsymbol{\mathsf{q}}^{*^T} \boldsymbol{\mathsf{R}}_{[\boldsymbol{\mathsf{q}}^*]}^{*^T}(\theta) \boldsymbol{\mathsf{A}}^{*} \boldsymbol{\mathsf{R}}_{[\boldsymbol{\mathsf{q}}^{*}]}^{*}(\theta) \boldsymbol{\mathsf{q}}^{*}. 
\end{equation}

\begin{figure}[H]
	\centering
		\includegraphics[width=0.9\textwidth]{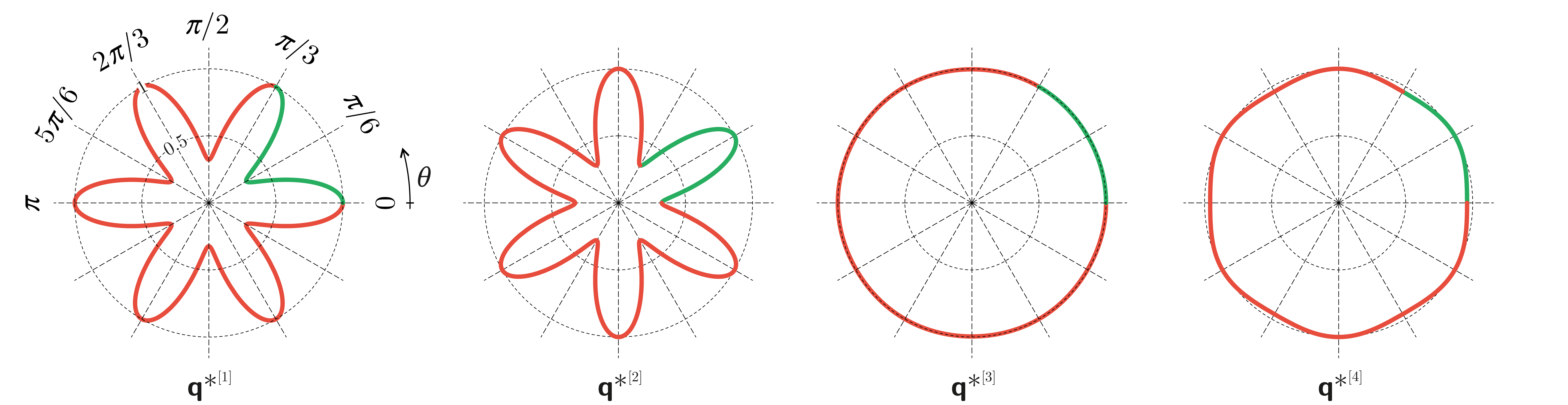}
	\caption{
	Polar diagrams of the curvature energy density $\mathcal{U}_{\boldsymbol{\mathsf{A}}^{*}}$ in a  $\mathsf{SGE}$ material (for the \lq condensed' representation),  equivalent to a hexagonal lattice with stiffness bar ratios $\widehat{k}/\overline{k}=10$, $\widetilde{k}/\overline{k}=100$. 
	The curvatures $\boldsymbol{\mathsf{q}}^{*[i]}$ ($i=1,...,4$), Eq. (\ref{eq:eigenvector}), are continuously rotated of an angle $\theta$ ranging between 0 and 2$\pi$. The diagrams show  the $\pi/3$ symmetry (green coloured curve part) of the constitutive response related to the matrix $\boldsymbol{\mathsf{A}}^{*}$.}
	\label{fig:PolarEne}
\end{figure}

Polar diagrams of the energy density $\mathcal{U}_{\boldsymbol{\mathsf{A}}^{*}}$ corresponding to a  second-gradient elastic material (in the \lq condensed' form), equivalent to a lattice with stiffness bar ratios $\widehat{k}/\overline{k}=10$, $\widetilde{k}/\overline{k}=100$,  is reported in Fig. \ref{fig:PolarEne} for a rotation $\theta$ ranging between 0 and 2$\pi$ of the following four curvature vectors,
\begin{equation}
\boldsymbol{\mathsf{q}}^{*[1]}=
\left[0,0,-0.278,0.961\right],\,
\boldsymbol{\mathsf{q}}^{*[2]}=
\left[0.967,-0.254,0,0\right],\,
\boldsymbol{\mathsf{q}}^{*[3]}=
\left[0,0,0.961,0.278\right],\,
\boldsymbol{\mathsf{q}}^{*[4]}=
\left[0.254,0.967,0,0\right],
\label{eq:eigenvector}
\end{equation}
corresponding to the four orthonormal eigenvectors of the \lq condensed' matrix $\boldsymbol{\mathsf{A}}^{*}$.
The polar diagrams, in which the values of the elastic energy density are divided by the maximum value attained during the 2$\pi$ rotation, highlight the direction sensitivity and symmetry of the material response with rotations, corresponding to $\theta= j\pi/3$ ($j\in\mathbb{Z}$).

The total elastic energy density $\mathcal{U}_{\mathsf{SGE}}$, Eq.~(\ref{eq:EneCont2}), given by the sum of the Cauchy, $\mathcal{U}_{\boldsymbol{\mathsf{C}}}$, curvature, $\mathcal{U}_{\boldsymbol{\mathsf{A}}^{*}}$, and mutual $\mathcal{U}_{\boldsymbol{\mathsf{M}}^{*}}$  terms,  can be rewritten as 
\begin{equation}
\mathcal{U}_{\mathsf{SGE}}=\frac{1}{2} \boldsymbol{\mathsf{t}}^{*^T} \boldsymbol{\mathsf{L}}^{*}\left(\overline{k},\widehat{k},\widetilde{k}\right)\boldsymbol{\mathsf{t}}^{*} ,
\label{eq:ene defi_posi}
\end{equation}
where the vector $\boldsymbol{\mathsf{t}}^{*}$ collects the strain and `condensed' curvature vectors and the matrix $\boldsymbol{\mathsf{L}}^{*}\left(\overline{k},\widehat{k},\widetilde{k}\right)$ collects the  matrices ($\boldsymbol{\mathsf{C}}$, $\boldsymbol{\mathsf{M}}^{*}$, $\boldsymbol{\mathsf{A}}^{*}$) characterizing the  \lq condensed' equivalent $\mathsf{SGE}$, so that
\begin{equation}
\boldsymbol{\mathsf{t}}^{*}=
\begin{bmatrix}
\boldsymbol{\mathsf{p}}^{\mathsf{SGE}^T}\\ 
\boldsymbol{\mathsf{q}}^{*^T}
\end{bmatrix},
\qquad
\boldsymbol{\mathsf{L}}^{*}=
\begin{bmatrix}
\boldsymbol{\mathsf{C}} & \boldsymbol{\mathsf{M}}^{*} \\ 
\boldsymbol{\mathsf{M}}^{*^T} & \boldsymbol{\mathsf{A}}^{*}.
\end{bmatrix}
\label{eq:vector_t_and_L}
\end{equation}
With the purpose of analyzing the directional property  of the total elastic energy, the following function is introduced 
\begin{equation}
\mathcal{L}^{*}=\boldsymbol{\mathsf{t}}^{*^T}\boldsymbol{\mathsf{L}}^{*}~\boldsymbol{\mathsf{t}}^{*} - \psi \left(\boldsymbol{\mathsf{t}}^{*} \scalp \boldsymbol{\mathsf{t}}^{*} - 1\right),
\label{eq:lagrangian}
\end{equation}
where $\psi$ is a Lagrangian multiplier. 
Imposing the stationarity of function (\ref{eq:lagrangian}) is equivalent to the following system of equations
\begin{equation}
\left\{
\begin{array}{l}
	\dfrac{\partial \mathcal{L}}{\partial \boldsymbol{\mathsf{t}}^{*}}=2 \left(\boldsymbol{\mathsf{L}}^{*} - \psi \bI \right) \boldsymbol{\mathsf{t}}^{*}=\b0,\\ [5mm]
	\dfrac{\partial \mathcal{L}}{\partial \psi}=\boldsymbol{\mathsf{t}}^{*} \scalp \boldsymbol{\mathsf{t}}^{*} - 1=\b0,
\end{array}\right.
\label{eq:stazio_lagrangian}
\end{equation}
with Eq. (\ref{eq:stazio_lagrangian})$_1$ defining the eigenvectors of the matrix $\boldsymbol{\mathsf{L}}^{*}$. These eigenvectors correspond to the seven principal deformations, while Eq. (\ref{eq:stazio_lagrangian})$_2$ constrains such eigenvectors to lie on the unitary hypersphere. For the $\mathsf{SGE}$ material (in the \lq condensed' form) equivalent to the lattice with stiffness ratios $\widehat{k}/\overline{k}=10$, $\widetilde{k}/\overline{k}=100$, the following seven orthonormal eigenvectors are obtained
%
%
%
%
%
%
\begin{equation}
\begin{array}{c}
\boldsymbol{\mathsf{t}}^{*[1]}=
\left[0.671,0.671,0,0,0,-0.091,0.303\right],
\boldsymbol{\mathsf{t}}^{*[2]}=
\left[-0.223,-0.223,0,0,0,-0.225,0.922\right],
\\[3mm]
\boldsymbol{\mathsf{t}}^{*[3]}=
\left[-0.011,-0.011,0,0,0,-0.970,-0.242\right],
\boldsymbol{\mathsf{t}}^{*[4]}=
\left[0,0,0,0.967,-0.254,0,0\right],
\\[3mm]
\boldsymbol{\mathsf{t}}^{*[5]}=
\left[0,0,0,0.254,0.967,0,0\right],
\boldsymbol{\mathsf{t}}^{*[6]}=
\left[0.707,-0.707,0,0,0,0,0\right],
\boldsymbol{\mathsf{t}}^{*[7]}=
\left[0,0,1,0,0,0,0\right],
\end{array}
\label{eq:eigenvector2}
\end{equation}
showing that only the first three eigenvectors have non-null components in both the strain and the curvature, while $\boldsymbol{\mathsf{t}}^{*[4]}$ and $\boldsymbol{\mathsf{t}}^{*[5]}$ have non-null components only in the curvature, and $\boldsymbol{\mathsf{t}}^{*[6]}$ and $\boldsymbol{\mathsf{t}}^{*[7]}$ only in the strain.
Note that the non-null curvature components of the vectors $\boldsymbol{\mathsf{t}}^{*[4]}$ and $\boldsymbol{\mathsf{t}}^{*[5]}$ coincide with the components of vectors $\boldsymbol{\mathsf{q}}^{*[2]}$ and $\boldsymbol{\mathsf{q}}^{*[4]}$, respectively.

\begin{figure}[H]
	\centering
	\centering
	\includegraphics[width=0.9\textwidth]{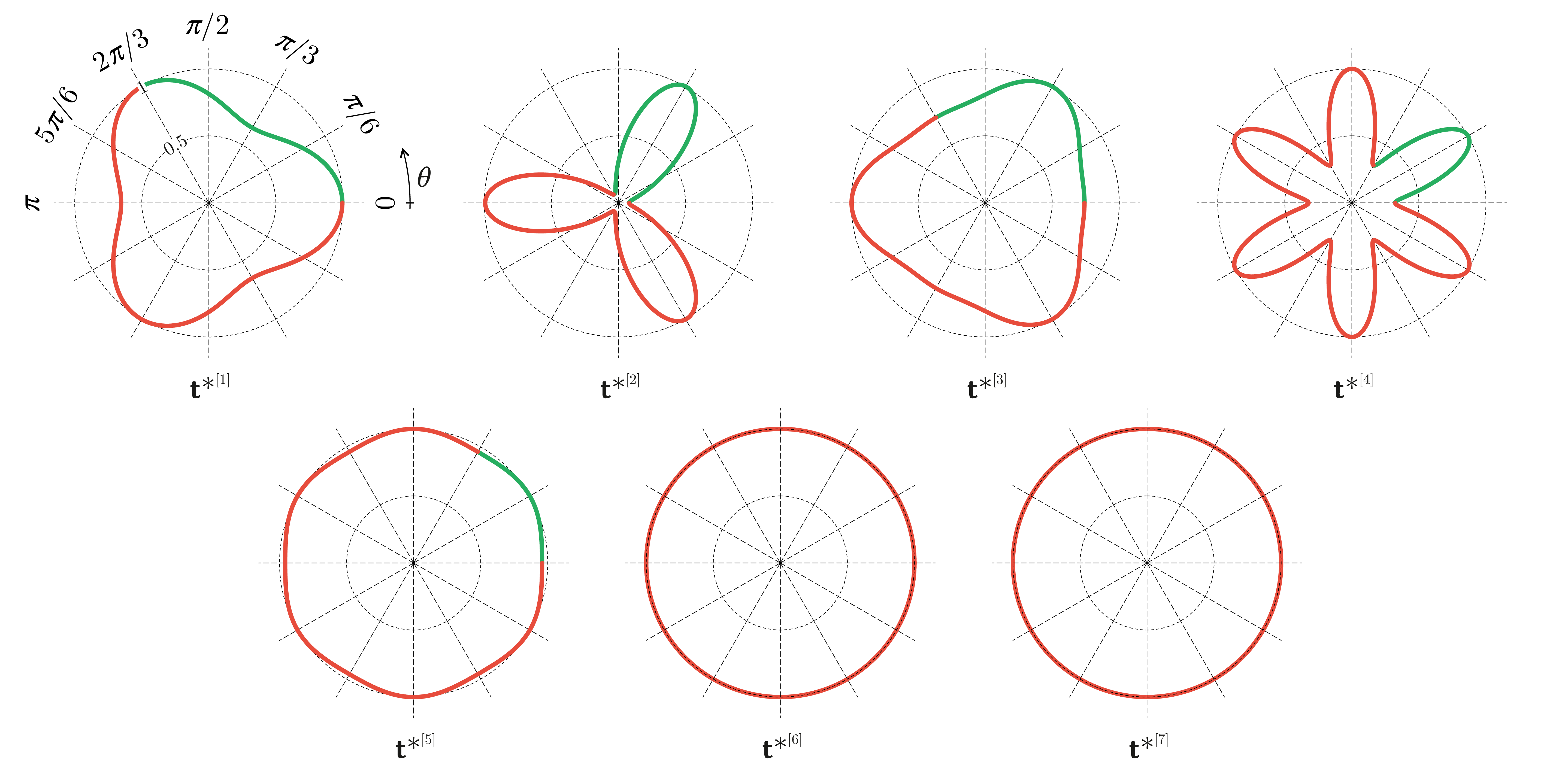}
	\caption{As for Fig. \ref{fig:PolarEne}, but for the total energy density $\mathcal{U}_{\mathsf{SGE}^{*}}$. Eigenvectors $\boldsymbol{\mathsf{t}}^{*[i]}$ ($i=1,...,7$), Eq. (\ref{eq:eigenvector2}), are continuously rotated of an angle $\theta$ ranging between 0 and 2$\pi$.
The diagrams show different symmetries:
(i.)
a $2\pi/3$ symmetry (when $\boldsymbol{\mathsf{t}}^{*[1]}$, $\boldsymbol{\mathsf{t}}^{*[2]}$, and $\boldsymbol{\mathsf{t}}^{*[3]}$ are rotated) related only to the constitutive matrix $\boldsymbol{\mathsf{M}}^*$ 
(ii.) 
a $\pi/3$ symmetry (when $\boldsymbol{\mathsf{t}}^{*[4]}$ and $\boldsymbol{\mathsf{t}}^{*[5]}$ are rotated) related only to the constitutive matrix $\boldsymbol{\mathsf{A}}^*$, 
(iii.) isotropy (when $\boldsymbol{\mathsf{t}}^{*[6]}$ and $\boldsymbol{\mathsf{t}}^{*[7]}$ are rotated) related only to the constitutive matrix $\boldsymbol{\mathsf{C}}$.}
	\label{fig:PolarEne2}
\end{figure}

The seven eigenvectors (\ref{eq:eigenvector2}) are rotated of an angle $\theta \in [0, 2\pi]$ to analyze the directional properties of the energy density (for a material equivalent to a lattice with stiffness ratios, $\widehat{k}/\overline{k}=10$, $\widetilde{k}/\overline{k}=100$), plotted as polar diagrams in Fig. \ref{fig:PolarEne2}.
The figure shows that: (i.) the constitutive matrix $\boldsymbol{\mathsf{M}}^*$ displays a $2\pi/3$--symmetry revealed by the rotation of the eigenvectors $\boldsymbol{\mathsf{t}}^{*[1]}$, $\boldsymbol{\mathsf{t}}^{*[2]}$, and $\boldsymbol{\mathsf{t}}^{*[3]}$, (ii.) 
the constitutive matrix $\boldsymbol{\mathsf{A}}^*$ displays a $\pi/3$--symmetry revealed by the rotation of the eigenvectors $\boldsymbol{\mathsf{t}}^{*[4]}$ and $\boldsymbol{\mathsf{t}}^{*[5]}$, and (iii.) 
the constitutive matrix $\boldsymbol{\mathsf{C}}$ is isoropic revealed by the rotation of the eigenvectors $\boldsymbol{\mathsf{t}}^{*[6]}$ and $\boldsymbol{\mathsf{t}}^{*[7]}$.  

\subsection{Positive definiteness of the constitutive tensors (in \lq condensed' form)}

A-priori stability of the response for a higher-order material equivalent to the lattice is related to the positive definiteness of the elastic energy.
It is known however that positive definiteness of higher-order solids equivalent to heterogeneous Cauchy materials is not always verified \cite{bacca2013mindlin, mattia2013mindlin,bigoni2007analytical, WHEEL201584}, so that the analysis of positive definiteness of the constitutive tensors so far derived is an important issue. 

\begin{figure}[H]
	\centering
	\includegraphics[width=0.5\textwidth]{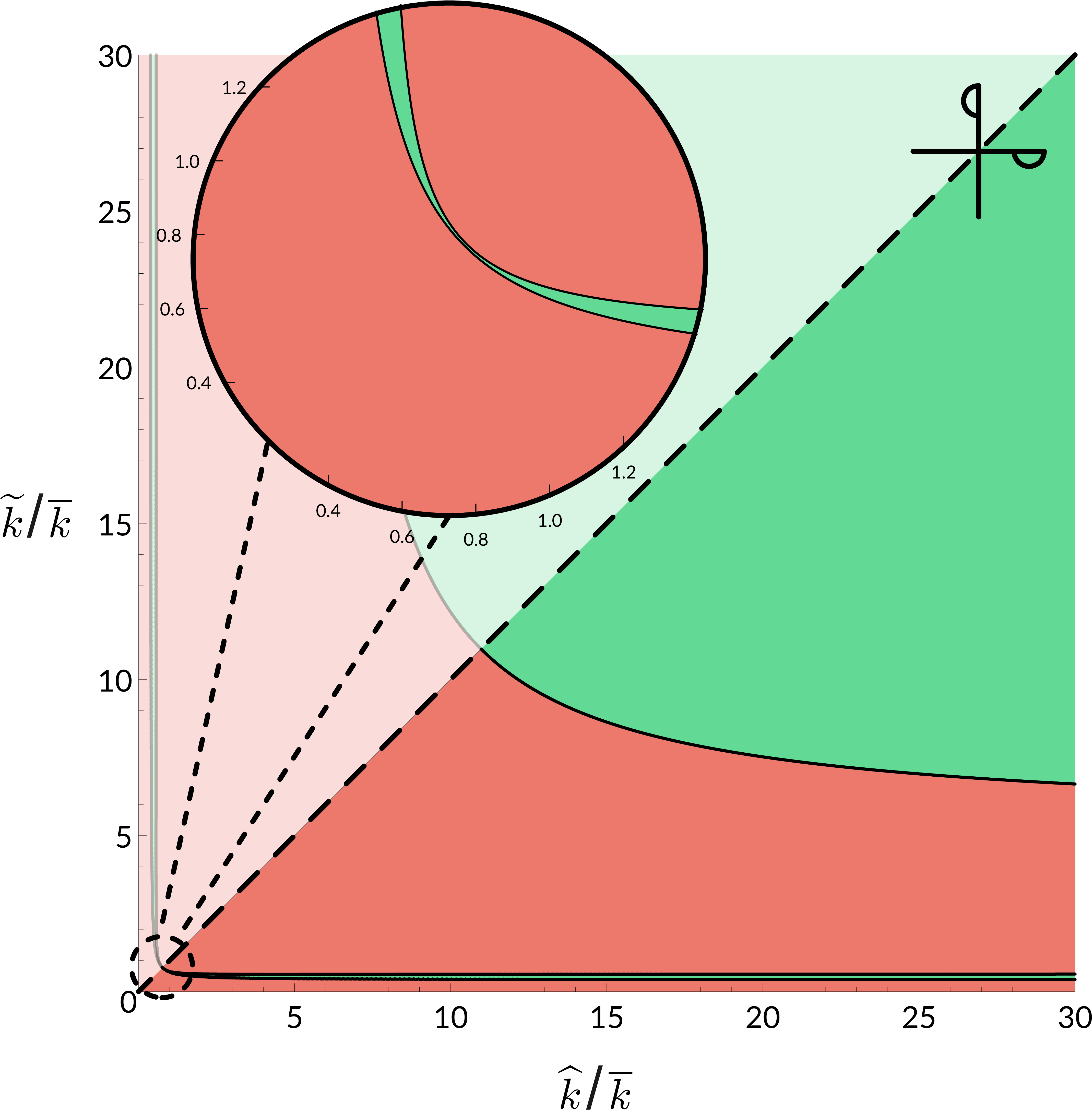}
	\caption{Regions in the bars' stiffness ($\widehat{k}/\overline{k}$--$\widetilde{k}/\overline{k}$) space corresponding to positive definite (green colour) or indefinite (red colour) equivalent second-gradient elastic material. Note the symmetry with respect to a line inclined at $\pi/4$ and that the case of equal stiffnessess, $\widehat{k}/\overline{k}$=$\widetilde{k}/\overline{k}$=1, corresponds to an indefinite material.}
	\label{fig:posi_def_Tot}
\end{figure}

With reference to the energy density $\mathcal{U}_{\mathsf{SGE}}$, Eq.~(\ref{eq:ene defi_posi}), the positive definiteness of the \lq condensed' $\mathsf{SGE}$ corresponds to the positive definiteness of the matrix $\boldsymbol{\mathsf{L}}^{*}$  collecting the  matrices $\boldsymbol{\mathsf{C}}$, $\boldsymbol{\mathsf{M}}^{*}$, and $\boldsymbol{\mathsf{A}}^{*}$, Eq. (\ref{eq:vector_t_and_L})$_2$.
Considering positive bar stiffnesses (for which the equivalent Cauchy material, Eq.~(\ref{eq:effettivo}), is positive definite), the sets of bar stiffnesses (made dimensionless through division by $\overline{k}$) for which the equivalent material is positive definite or indefinite are displayed in Fig. \ref{fig:posi_def_Tot} as green and red regions, respectively.

The two cases of isotropy, given by Eq.~(\ref{eq:Isotropy}), are remarkable. In fact, for both of them the equivalent material is positive definite, while in the case of three equal bar stiffnesses, the equivalent material, not isotropic at higher-order, is also not positive definite.
Note that the positive definiteness of the equivalent material is not affected by a permutation of $\widehat{k}$ with $\widetilde{k}$, so that the symmetry line inclined at $\pi/4$ is observed in the figure. The non positive definiteness issue is discussed further by analyzing the elastic energy stored within a circular domain at increasing its size.

\paragraph{Variation of the elastic energy with the characteristic length.}
Even in the case of non positive definite equivalent constitutive response, the elastic energy stored within a disk of equivalent material of radius $\rho$ centred at the origin of the reference system and generated by a quadratic displacement field (producing an equilibrated stress field) is always positive whenever the ratio between the disk radius $\rho$ and the hexagonal lattice side $\ell$ exceeds a finite value.
Restricting attention to a purely quadratic displacement conditions ($\balpha=\b0$)
\footnote{A purely quadratic displacement ($\balpha=\b0$) is a strongly demanding condition for positive definiteness of the elastic energy within a finite domain, because the  Cauchy part $\mathsf{U}_{\boldsymbol{\mathsf{C}}}$ of the elastic energy, always positive, is only partially (through tensor $\bbeta$) activated.},
the integration of the energy density, Eq.~(\ref{eq:EneCont2}), over the aforementioned disk yields
\begin{equation}
\mathsf{U}_{\mathsf{SGE}} = 
\underbrace{\rho^4 \mathcal{C} \left(\boldsymbol{\mathsf{C}}\left(\overline{k},~\widehat{k},~\widetilde{k}\right),\bbeta\right) }_{\mathsf{U}_{\boldsymbol{\mathsf{C}}}}   + 
\underbrace{ \rho^2 \ell^2 \,\mathcal{A}\left(\boldsymbol{\mathsf{A}}^{*}\left(\overline{k},~\widehat{k},~\widetilde{k}\right),\bbeta\right)}_{\mathsf{U}_{\boldsymbol{\mathsf{A}}^{*}}},
\label{eq:ene_varia_chara_leng}
\end{equation}
where $\mathcal{C} \left(\boldsymbol{\mathsf{C}}(\overline{k},~\widehat{k},~\widetilde{k}),\bbeta\right)$ is always positive, while $\mathcal{A}\left(\boldsymbol{\mathsf{A}}^{*}(\overline{k},~\widehat{k},~\widetilde{k}),\bbeta\right)$ has no sign restriction.

Eq. (\ref{eq:ene_varia_chara_leng}) shows that the Cauchy $\mathsf{U}_{\boldsymbol{\mathsf{C}}}$ and the higher-order $\mathsf{U}_{\boldsymbol{\mathsf{A}}^{*}}$ energies are ruled by different powers in the disk radius $\rho$, so that they scale differently with it. On the other hand, the mutual energy $\mathsf{U}_{\boldsymbol{\mathsf{M}}^{*}}$ is null, because it follows from an integration of an odd function over a symmetric domain.

Introducing the parameter $\mathcal{R}$ as the ratio between the $\mathsf{U}_{\boldsymbol{\mathsf{A}}^{*}}$ and $\mathsf{U}_{\boldsymbol{\mathsf{C}}}$ 
\begin{equation}
	\mathcal{R} = \frac{\mathsf{U}_{\boldsymbol{\mathsf{A}}^{*}}}{\mathsf{U}_{\boldsymbol{\mathsf{C}}}},
	\label{errerre}
\end{equation} 
the elastic energy $\mathsf{U}_{\mathsf{SGE}}$ (\ref{eq:EneCont2})
 can be rewritten as
 \begin{equation}
\mathsf{U}_{\mathsf{SGE}} = (1+\mathcal{R} )\mathsf{U}_{\boldsymbol{\mathsf{C}}}.
\end{equation}
Considering expression (\ref{eq:ene_varia_chara_leng}), the parameter $\mathcal{R}$  (\ref{errerre}) reduces to
\begin{equation}\label{errerrerre}
	\mathcal{R} = \frac{\mathcal{A}\left(\boldsymbol{\mathsf{A}}^{*}\left(\overline{k},~\widehat{k},~\widetilde{k}\right),\bbeta\right)}{\mathcal{C} \left(\boldsymbol{\mathsf{C}}\left(\overline{k},~\widehat{k},~\widetilde{k}\right),\bbeta\right)} \left(\frac{\ell}{\rho}\right)^2,
\end{equation} 
and the following two considerations can be made:
\begin{itemize}
	\item a positive value for the stored energy $\mathsf{U}_{\mathsf{SGE}}$, Eq.~(\ref{eq:ene_varia_chara_leng}), is achieved when the following constraint on $\mathcal{R}$ is enforced
	\begin{equation} 
	\frac{\mathcal{A}\left(\boldsymbol{\mathsf{A}}^{*}\left(\overline{k},~\widehat{k},~\widetilde{k}\right),\bbeta\right)}{\mathcal{C} \left(\boldsymbol{\mathsf{C}}\left(\overline{k},~\widehat{k},~\widetilde{k}\right),\bbeta\right)} \left(\frac{\ell}{\rho}\right)^2> - 1 .
	\label{eq:constrain_ene_posi}
	\end{equation}
	which is always satisfied if 
	 $\mathcal{A}\left(\boldsymbol{\mathsf{A}}^{*}\left(\overline{k},~\widehat{k},~\widetilde{k}\right),\bbeta\right)$
	 is greater than zero, since  $\mathcal{C} \left(\boldsymbol{\mathsf{C}}\left(\overline{k},~\widehat{k},~\widetilde{k}\right),\bbeta\right)$ always is;
	
	\item the parameter  $\mathcal{R}$ increases when the higher-order part $\mathsf{U}_{\boldsymbol{\mathsf{A}}^{*}}$ of the elastic energy is increased with respect to the Cauchy $\mathsf{U}_{\boldsymbol{\mathsf{C}}}$ energy. A large value of the parameter 
	$\mathcal{R}$ can always be achieved by increasing either the ratio $\ell/\rho$ or the ratio $\left.\mathcal{A}\left(\boldsymbol{\mathsf{A}}^{*}\left(\overline{k},~\widehat{k},~\widetilde{k}\right),\bbeta\right)\right/\mathcal{C} \left(\boldsymbol{\mathsf{C}}\left(\overline{k},~\widehat{k},~\widetilde{k}\right),\bbeta\right)$. 
\end{itemize}

\begin{figure}[H]
	\centering
	\includegraphics[width=0.8\textwidth]{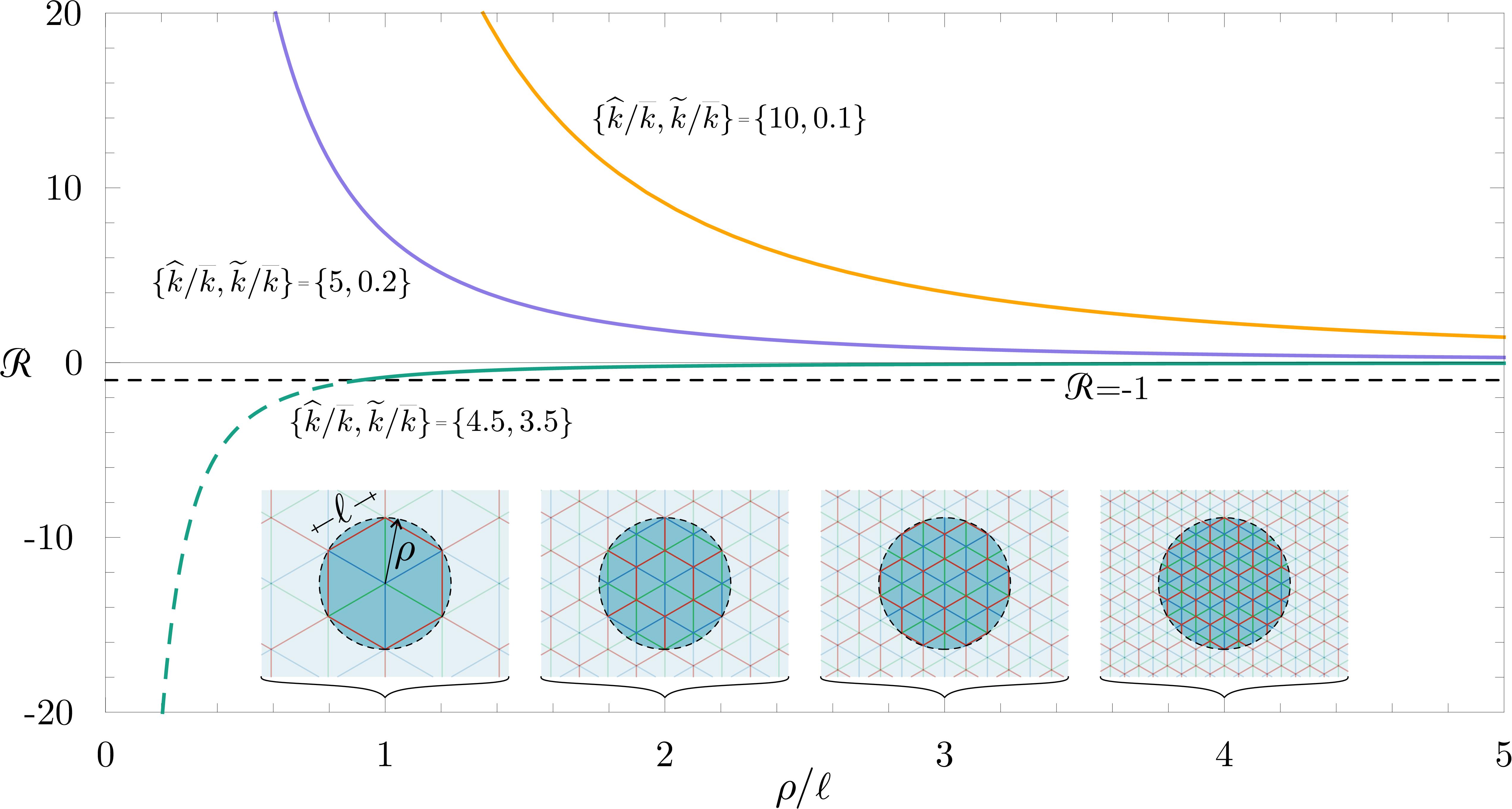}
	\caption{Parameter $\mathcal{R}$, Eq. (\ref{errerre}), as a function of the ratio $\rho/\ell$, for $\beta_{222}=-\beta_{112}=-\beta_{211}=1$. Different  pairs of bars' stiffness are reported $\{\widehat{k}/\overline{k},\widetilde{k}/\overline{k}\}$, given by $\left\{4.5,3.5\right\}$~(green), $\left\{5,0.2\right\}$~(purple) and $\left\{10,0.1\right\}$~(orange). Note that the ratio is always positive at sufficiently large $\rho/\ell$.}
	\label{fig:EneNegaMini}
\end{figure}

The behaviour of parameter $\mathcal{R}$ as a function of the ratio $
\ell/\rho$ is reported in Fig. \ref{fig:EneNegaMini}, for three different values of the pairs of bars' stiffness  $\{\widehat{k}/\overline{k},\widetilde{k}/\overline{k}\}$ and for the following values of the non-null components of $\bbeta$: $\beta_{222}=-\beta_{112}=-\beta_{211}=1$. 
The figure shows that the stored energy becomes always positive when the disk radius $\rho$ is sufficiently large.

As a closing remark, it is observed  that the higher-order energy  contribution $\mathsf{U}_{\boldsymbol{\mathsf{A}}^{*}}$ becomes negligible with respect to the Cauchy part $\mathsf{U}_{\boldsymbol{\mathsf{C}}}$ at a threshold value of the ratio $\rho/\ell$, function of the bars' stiffnesses.

\subsection{The constitutive law for the equivalent second-gradient elastic material}
\label{Sec:relaxed}

Having used for the identification (performed in Part I of this study) a quadratic displacement field which produces an equilibrated stress field, 
an equivalent second-gradient material has been so far defined in a \lq condensed' form, corresponding to a class of $\infty^4$ second-gradient materials, all providing a  correct energy matching with the periodic planar lattice. 
At this stage, a \lq relaxation of the constraints' has to be introduced to yield an equivalent second-gradient elastic material in a, say, `standard form'.
This relaxation can be introduced in several ways, as for example exploiting an optimization scheme. However, in the following a simple approach is pursued, which corresponds 
to imposing (still at the first indentification step)  an equivalence between the elastic energies of the lattice and the equivalent solid in which the equilibrium constraint on the coefficients $\beta_{ijk}$ (Eq. (46) and (75) in \cite{rizzipt1}) is neglected, therefore removing the concept itself of \lq condensed' material. 

Removing the equilibrium constraint yields the definition of a unique second-gradient elastic material, through the following expressions for the four coefficients $\mathsf{Q}_{35}$, $\mathsf{A}_{11}$, $\Delta\mathsf{M}_{15}$, and $\Delta\mathsf{M}_{16}$ 
\begin{equation}
\begin{split}
	\mathsf{Q}_{35}=&
		\dfrac{I_{[1]}}{4 \mathcal{E}} \left[ 26 \overline{k}^3 \left(\widehat{k}+\widetilde{k}\right)^2-2 \overline{k}^2 \left(\widehat{k}+\widetilde{k}\right) \left(5 \widehat{k}^2-16 \widehat{k} \widetilde{k}+5 \widetilde{k}^2\right) + \right.\\[3mm]
		&\left. I_{[3]} \left(-47 \widehat{k}^2+40 \widehat{k} \widetilde{k}-47 \widetilde{k}^2\right)-10\widehat{k}^2 \widetilde{k}^2 \left(\widehat{k}+\widetilde{k}\right)\right],
	\\[3mm]
	\Delta \mathsf{A}_{11}=&
		-\dfrac{\left(2 \overline{k}^2 \left(\widehat{k}+\widetilde{k}\right)+\overline{k} \left(2 \widehat{k}^2+15 \widehat{k} \widetilde{k}+2 \widetilde{k}^2\right)+2 \widehat{k} \widetilde{k} \left(\widehat{k}+\widetilde{k}\right)\right)^2 \ell^2}{192 \sqrt{3} I_{[2]}^2 \mathcal{E}}
		\left[-60 \overline{k}^4\left(\widehat{k}+\widetilde{k}\right)\right.\\[3mm]
		&+\overline{k}^3 \left(-155 \widehat{k}^2+818 \widehat{k}\widetilde{k}-155 \widetilde{k}^2\right) - \overline{k}^2 \left(\widehat{k}+\widetilde{k}\right)\left(35 \widehat{k}^2-211 \widehat{k} \widetilde{k}+35 \widetilde{k}^2\right)\\[3mm]
		&\left.+ I_{[3]} \left(182 \widehat{k}^2-253 \widehat{k} \widetilde{k}+182 \widetilde{k}^2\right)+28 \widehat{k}^2 \widetilde{k}^2\left(\widehat{k}+\widetilde{k}\right)\right],
	\\[3mm]
	\Delta\mathsf{M}_{15} =& \Delta\mathsf{M}_{16} = 0.
\end{split}
\label{eq:coeff_q35}
\end{equation}
where $\mathcal{E}$ is defined as
\begin{equation}
\begin{split}
	\mathcal{E} =&
		 5 \overline{k}^4 \left(\widehat{k}+\widetilde{k}\right)^2+\overline{k}^3 \left(\widehat{k}+\widetilde{k}\right) \left(10 \widehat{k}^2-69 \widehat{k} \widetilde{k}+10 \widetilde{k}^2\right)+\overline{k}^2 \left(5 \widehat{k}^4+13 \widehat{k}^3 \widetilde{k}-78 \widehat{k}^2\widetilde{k}^2+13 \widehat{k} \widetilde{k}^3+5 \widetilde{k}^4\right) \\[3mm]
		 &-13 I_{[3]}\left(\widehat{k}+\widetilde{k}\right) \left(2 \widehat{k}^2-3 \widehat{k} \widetilde{k}+2 \widetilde{k}^2\right)-4 \widehat{k}^2 \widetilde{k}^2 \left(\widehat{k}+\widetilde{k}\right)^2.
\end{split}
\end{equation}
A substitution of the above coefficients in the matrices (\ref{eq:Q_DM_DA}), which are also constrained by the material symmetry of the \lq condensed' solid remarked in section \ref{sec:Aniso}, yields the two constitutive tensors $\boldsymbol{\mathsf{M}}\left(\overline{k}, \widehat{k}, \widetilde{k}\right)$ and $\boldsymbol{\mathsf{A}}\left(\overline{k}, \widehat{k}, \widetilde{k}\right)$ anticipated with Eqn.~(\ref{eq:effettivo}).
It is highlighted that the adopted choice for the coefficients in Eqs.~(\ref{eq:coeff_q35}) provides a domain of positive definiteness identical for both the \lq condensed' (Fig. \ref{fig:posi_def_Tot}) and the extended forms of the equivalent material.

\section{Validation of the identified $\mathsf{SGE}$ material}

\begin{figure}[!htb]
	\centering
	\includegraphics[width=\textwidth]{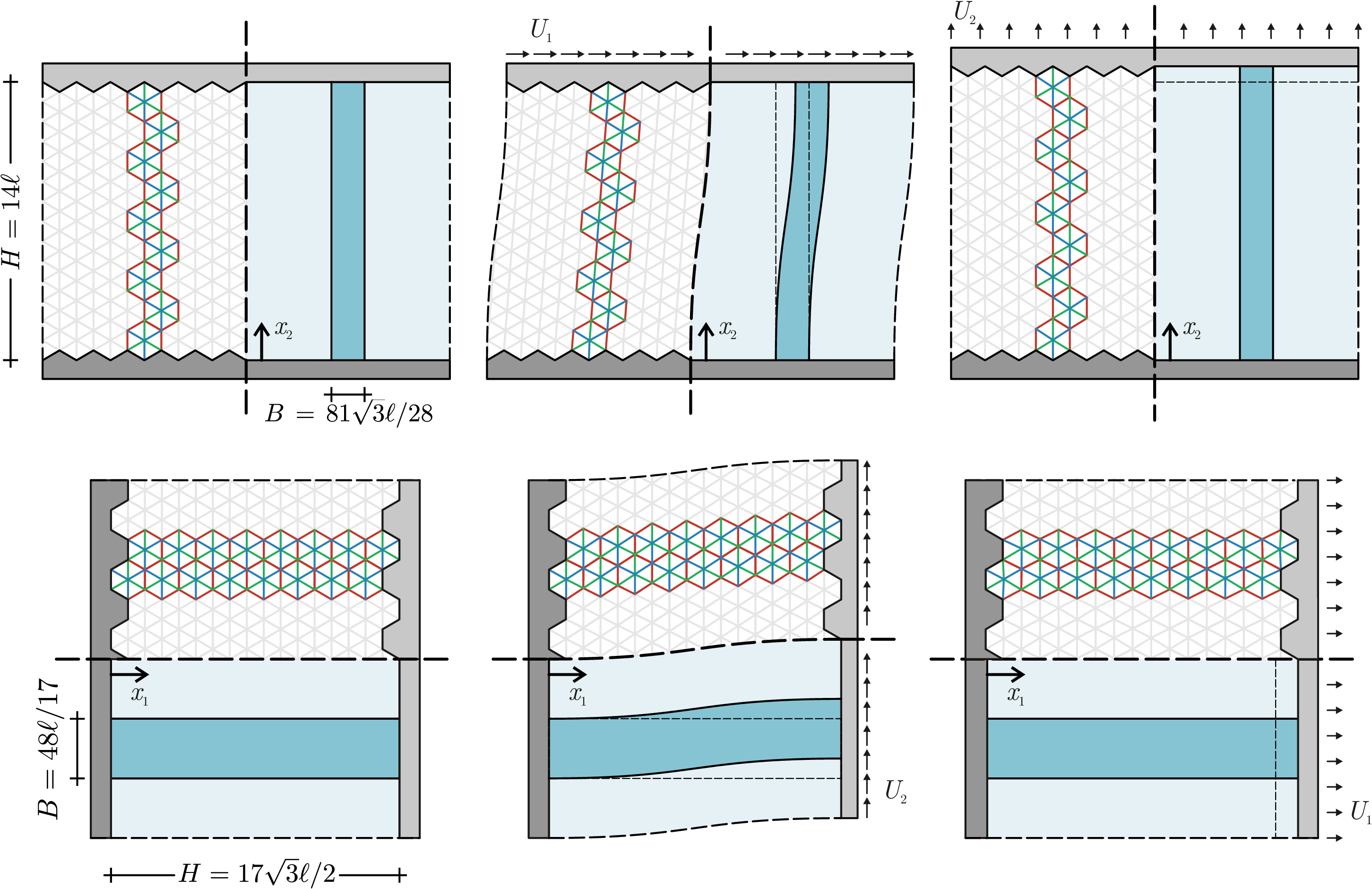}
	\caption{Upper row from left to right: 
		Undeformed lattice and equivalent $\mathsf{SGE}$  material; deformed configuration induced by a simple shear parallel to the horizontal direction; deformed configuration induced by a uniaxial strain in the vertical direction. 
		Lower row from left to right: 
		Undeformed lattice and equivalent $\mathsf{SGE}$  material; deformed configuration induced by a simple shear parallel to the vertical direction; deformed configuration induced by a uniaxial strain in the horizontal direction.
		The deformed configurations (which are exact) for the lattice are displayed for bars having the same stiffness, $\widehat{k}/\overline{k}=\widetilde{k}/\overline{k}=1$, which corresponds to centrosymmetric response.}
	\label{fig:shearx1}
\end{figure}

The second-gradient elastic material previously obtained to be equivalent to the hexagonal lattice structure is validated through comparisons between their corresponding response to the same mechanical input. In particular, two basic boundary value problems are analyzed, namely, simple shear and uniaxial strain, providing the four deformations shown in Fig. \ref{fig:shearx1}, which have been reported after an analytical determination (so that they are exact). In all of these cases the response of the equivalent homogeneous material is provided in closed-form expressions, while that of the lattice is obtained by solving a linear system. 

In the absence of body forces, the stress $\sigma_{ij}$ and the double-stress $\tau_{kij}$ fields within a homogeneous $\mathsf{SGE}$ material are governed by the equilibrium equations 
\begin{equation}
\sigma_{ij,j}-\tau_{kji,kj} =  0,
\label{eq:eq_equa_gen}
\end{equation}
moreover, the components of the resultant traction vector $\bP\left(\bn\right)$ associated with the surface of normal unit vector $\bn$ are given by
\begin{equation}
P_k(\bn)=\sigma_{jk}~n_{j} - n_{i}~n_{j}~D\left(\tau_{ijk}\right)-2n_{j}~D_{i}\left(\tau_{ijk}\right) + \left(n_{i}~n_{j}~D_{l}\left(n_{l}\right)-D_{j}\left(n_{i}\right)\right)\tau_{ijk}, 
\label{eq:eq_equa_gen2}
\end{equation}
where the differential operators $D_{j}\left(\scalp \right)$
and $D \left(\scalp \right)$ are defined as 
\begin{equation}
D_{j}\left(\scalp \right) =\left(\delta_{jl}-n_{j}n_{l}\right)\left(\scalp \right)_{,l},\qquad
D \left(\scalp \right)=n_{l} \left(\scalp \right)_{,l},
\end{equation}
with the comma defining differentiation with respect to the coordinates $x_l$. 

The mechanical responses of the lattice and of the related equivalent homogeneous solid are compared for  different stiffness ratios pairs $\widehat{k}/\overline{k}$ and $\widehat{k}/\overline{k}$  and corresponding constitutive parameters  $\mu$, $\lambda$, $\mathsf{m}$, $\mathsf{a}_{11}$, $\mathsf{a}_{22}$, $\mathsf{a}_{33}$, and $\mathsf{a}_{44}$  of the $\mathsf{SGE}$. All of these parameters are listed for  the six  Example cases (Ex1, Ex2, Ex3, Ex4, Ex5, and Ex6) in Table~\ref{tab:stiff_combo}. The comparison is performed by analysing displacement and strain energy density fields.
As far as regards the geometry:
\begin{itemize}
\item for simple shear and uniaxial strain respectively aligned parallel to the $x_1$--axis and to the $x_2$--axis, the lattice structure is considered to be realized as the periodic repetition along the $x_1$--axis of one line of 9 unit cells, so that  the strip of equivalent homogeneous solid has sides $H=14 \ell$ and (by imposing the equivalence of area) $B=81\sqrt{3}\ell/28$, Fig. \ref{fig:shearx1} (top);
\item for simple shear and uniaxial strain respectively aligned parallel to the $x_2$--axis and to the $x_1$--axis, the periodic repetition along the $x_2$ axis of two lines of 8 unit cells realizes the lattice structure, so that the equivalent homogeneous strip 
has sides $H=17\sqrt{3}\ell/2 $ and (by imposing the equivalence of area) $B=48 \ell/17$, Fig. \ref{fig:shearx1} (bottom). 
\end{itemize}

\begin{table}[H]
	\centering
	\caption{The stiffness ratios $\widehat{k}/\overline{k}$ and $\widehat{k}/\overline{k}$ and the corresponding equivalent constitutive parameters $\mu$, $\lambda$, $\mathsf{m}$, $\mathsf{a}_{11}$, $\mathsf{a}_{22}$, $\mathsf{a}_{33}$, and $\mathsf{a}_{44}$ (made dimensionless through division by the bar stiffness $\overline{k}$) used to compare the solutions of the boundary value problems shown in Figs.~\ref{fig:Shear_x_A}-\ref{fig:Uniaxial_x_NCS}.}
	\label{tab:stiff_combo}
	\resizebox{0.7\textwidth}{!}{$
		\centering
		\renewcommand{\arraystretch}{1.5}
		\begin{tabular}{c|c|c|c|c|c|c|c|c|c|}
		\cline{2-10}
		& $\widehat{k}/\overline{k}$ & $\widetilde{k}/\overline{k}$ & $\mu /\overline{k}$ & $\lambda /\overline{k}$ & $\mathsf{m} /\overline{k}$ & $\mathsf{a}_{11} /\overline{k}$ & $\mathsf{a}_{22} /\overline{k}$ & $\mathsf{a}_{33} /\overline{k}$ & $\mathsf{a}_{44} /\overline{k}$ \\ \hline
		\multicolumn{1}{|c|}{Ex1} & 1     & 1     & 0.433 & 0.433 &  0    & 1.27  & 0.153			   & 0.135 & 1.25     \\ \hline
		\multicolumn{1}{|c|}{Ex2} & 25    & 25    & 1.20  & 13.52 & 0     & 5.902 & 0.426			   & 0.554 & 6.03 \\ \hline
		\multicolumn{1}{|c|}{Ex3} & 1500  & 10    & 1.18  & 435   & 78    & 129   & 0.413              & 14.4   & 143  \\ \hline
		\multicolumn{1}{|c|}{Ex4} & 0.75  & 0.75  & 0.354 & 0.367 & 0     & 1.19  & 0.125              & 0.127  & 1.188 \\ \hline
		\multicolumn{1}{|c|}{Ex5} & 0.05  & 0.5   & 0.056 & 0.39  & 0.06  & 0.823 & 5.45$\cdot 10^{-4}$  & 0.034  & 0.857 \\ \hline
		\multicolumn{1}{|c|}{Ex6} & 0.5  & 0.05   & 0.056 & 0.39  & -0.06  & 0.823 & 5.45$\cdot 10^{-4}$  & 0.034  & 0.857 \\ \hline
		\end{tabular}
		$}
\end{table}

\subsection{Simple shear problem}

The kinematical boundary conditions for a $\mathsf{SGE}$ material subject to a simple shear aligned parallel to the $x_I$--axis ($I=1,2$) are 
\begin{equation}
\label{BCsimpleshear}
	\begin{array}{lllll}
	& u_I\left(x_J=0\right)=0,
	&u_I\left(x_J=H\right)=U_I,
	&u_{I,J}\left(x_J=0\right)=0,
	& u_{I,J}\left(x_J=H\right)=0,
	\\[3mm]
	&u_J\left(x_J=0\right)=0,&u_J\left(x_J=H\right)=0,
	&u_{J,J}\left(x_J=0\right)=0,
	&u_{J,J}\left(x_J=H\right)=0,
	\end{array}
	\qquad
	\mbox{for}\left\{
	\begin{array}{llll}
	I,J=1,2,\\
	I\neq J,
	\end{array}
	\right.
\end{equation} 
where $U_I$ is the prescribed displacement aligned parallel to the $x_I$-direction and applied at $x_J=H$ (reported in the centre of Fig. \ref{fig:shearx1} for $I=1$ and $J=2$, top and bottom part, respectively). Because the boundary conditions are independent of $x_I$ and the $\mathsf{SGE}$ material is homogeneous, the displacement fields are only dependent on  $x_J$,
\begin{equation}
u_I=u_I(x_J),\qquad
u_J=u_J(x_J),
\end{equation}
so that the equilibrium equations~(\ref{eq:eq_equa_gen})  in the case of simple shear aligned parallel the $x_1$--axis ($I=1$, $J=2$) reduce to 
\begin{equation}
\label{eq:equilibrium_shear1}
\left\{
\begin{array}{lllll}
u_{1,22} (x_2) - \ell^2 \dfrac{\mathsf{a}_{22} }{ \mu} u_{1,2222}(x_2) = 0, \\[3mm]
u_{2,22} (x_2) - \ell^2 \dfrac{\mathsf{a}_{44} }{ (\lambda +2 \mu ) } u_{2,2222}(x_2) = 0,
\end{array}
\right.
\end{equation}
and the resultant traction components Eq.~(\ref{eq:eq_equa_gen2}) which are generated on a surface parallel to the $x_1$--axis (and therefore $\bn=\be_2$) result to be 
\begin{equation}
\label{eq:traction_x1}
\left\{
\begin{array}{lllll}
P_{1}\left(\be_{2}\right) = \sigma_{12} - \tau_{221,2} \,\, , \\[3mm]
P_{2}\left(\be_{2}\right) = \sigma_{22} - \tau_{222,2} \,\, .
\end{array}
\right.
\end{equation}

Differently, if the simple shear is aligned parallel to the $x_2$--axis ($I=2$, $J=1$), the equilibrium equations  (\ref{eq:eq_equa_gen}) become 
\begin{equation}
\label{eq:equilibrium_shear2}
\left\{
\begin{array}{llllll}
u_{1,11}(x_{1}) + \ell \left(\dfrac{\mathsf{m}}{\lambda +2 \mu } u_{2,111}(x_{1}) - \ell \dfrac{\mathsf{a}_{11}}{\lambda +2 \mu } u_{1,1111}(x_{1})\right) = 0, \\[3mm]
u_{2,11} (x_1) - \ell \left( \dfrac{\mathsf{m}}{\mu } u_{1,111}(x_1) + \ell \dfrac{\mathsf{a}_{33} }{\mu }u_{2,1111}(x_1)\right) = 0,
\end{array}
\right.
\end{equation}
and the resultant traction components Eq.~(\ref{eq:eq_equa_gen2}) which are generated on a surface parallel to the $x_2$--axis result to be
\begin{equation}
\label{eq:traction_x2}
\left\{
\begin{array}{lllll}
P_{1}\left(\be_{1}\right) = \sigma_{11} - \tau_{111,1} \,\, , \\[3mm]
P_{2}\left(\be_{1}\right) = \sigma_{12} - \tau_{112,1} \,\, ,
\end{array}
\right.
\end{equation}
where $\mathsf{a}_{22} = \mathsf{a}_{11}-2 (\mathsf{a}_{16}+\mathsf{a}_{26})$, $\mathsf{a}_{33} = \mathsf{a}_{11}+\mathsf{a}_{12}-2 (\mathsf{a}_{16}+\mathsf{a}_{26})-\mathsf{a}_{34}$, and $\mathsf{a}_{44} = \mathsf{a}_{11}+\mathsf{a}_{12}-\mathsf{a}_{34}$, in agreement with Eq.~(\ref{eq:effettivo})$_{3}$.

Note that the differential equations (\ref{eq:equilibrium_shear1}) and (\ref{eq:equilibrium_shear2}) are uncoupled when the  simple shear is aligned parallel to the $x_1$--axis, while these become coupled when the simple shear is aligned parallel to the $x_2$--axis.

The two cases of simple shear aligned parallel the $x_1$--axis and parallel to the $x_2$--axis are separately analyzed below.

\subsubsection{Simple shear parallel to the $x_1$--axis}

Solving the equilibrium equations (\ref{eq:equilibrium_shear1}) subject to the boundary conditions (\ref{BCsimpleshear}) leads to the following solution in terms of displacement fields within the $\mathsf{SGE}$ solid
\begin{equation}
u_{1}(x_2) =
\dfrac{ \cosh \left(\mathcal{Z}_{s}\right)x_2 +\left[\sinh\left(\mathcal{Z}_{s}\left(1-\dfrac{2 x_2}{H}\right)\right)-\sinh\left(\mathcal{Z}_{s}\right)\right]\dfrac{H}{2\mathcal{Z}_{s}}}{\cosh \mathcal{Z}_{s}-\dfrac{1}{\mathcal{Z}_{s}}  \sinh\mathcal{Z}_{s}}\dfrac{U_2}{H},
\quad
u_{2}(x_2) = 0,
\end{equation}
and in terms of stress and double stress fields
\begin{equation}
\begin{array}{ccccc}
\sigma_{12}(x_2) =\mu~
\dfrac{2 \sinh\left( \dfrac{x_2}{H} \mathcal{Z}_{s} \right) \sinh\left(\mathcal{Z}_{s}\left(1-\dfrac{x_2}{H}\right)\right)}{ \cosh\mathcal{Z}_{s}-\dfrac{1}{\mathcal{Z}_{s}}\sinh\mathcal{Z}_{s}}\dfrac{U_2}{H},
\quad
\tau_{221}(x_2) = \dfrac{\mathsf{a}_{22}}{\mathsf{a}_{12}} \tau_{111}(x_2), \\[7mm]
\tau_{111}(x_2) =\ell~\mathsf{a}_{12}~
\dfrac{ \dfrac{2\ell}{H} \mathcal{Z}_{s} \sinh\left(\mathcal{Z}_{s}\left(1-\dfrac{2x_2}{H}\right)\right) }{ \cosh\mathcal{Z}_{s}-\dfrac{1}{\mathcal{Z}_{s}}\sinh\mathcal{Z}_{s}}\dfrac{U_2}{H},
\quad
\tau_{122}(x_2) =  \dfrac{\mathsf{a}_{26}}{\mathsf{a}_{12}} \tau_{111}(x_2).
\end{array}
\end{equation}

Note that in the above-reported solution the parameter $\mathsf{m}$ does not appear, so that the considered boundary value problem does not activate the possible non-centrosymmetric response.

The components of the resultant traction vector associated with the normal $\bn=\be_2$ are found to be independent of $x_2$ in the form
\begin{equation}
P_{1}\left(\be_{2}\right) = \dfrac{\mu}{1 - \dfrac{1}{\mathcal{Z}_{s}}\tanh\mathcal{Z}_{s}}\dfrac{U_2}{H}, \qquad
P_{2}\left(\be_{2}\right) = 0,
\end{equation}
and the three contributions $\mathcal{U}_{\boldsymbol{\mathsf{C}}}$, $\mathcal{U}_{\boldsymbol{\mathsf{M}}}$, and $\mathcal{U}_{\boldsymbol{\mathsf{A}}}$ to the strain energy density are given by 
\begin{equation}
\begin{array}{ccccc}
\mathcal{U}_{\boldsymbol{\mathsf{C}}}(x_2)=
\mu \dfrac{2 \sinh^{2}\left(\mathcal{Z}_{s}\left(1-\dfrac{x_2}{H}\right)\right)\sinh^{2}\left(\dfrac{x_2}{H} \mathcal{Z}_{s} \right) }{\left( \cosh\mathcal{Z}_{s}-\dfrac{1}{\mathcal{Z}_{s}}\sinh\mathcal{Z}_{s}\right)^{2}} \dfrac{U_2^{2}}{H^{2}},
\quad
\mathcal{U}_{\boldsymbol{\mathsf{M}}}(x_2)=0, 
\\
\mathcal{U}_{\boldsymbol{\mathsf{A}}}(x_2)=
\mu \dfrac{ \sinh^{2}\left(\mathcal{Z}_{s}\left(1-\dfrac{2x_2}{H}\right)\right)}{2\left( \cosh\mathcal{Z}_{s}-\dfrac{1}{\mathcal{Z}_{s}}\sinh\mathcal{Z}_{s}\right)^{2}} \dfrac{U_2^{2}}{H^{2}},
\end{array}
\end{equation}
where
\begin{equation}
\mathcal{Z}_{s}=\dfrac{H }{2 \ell}\sqrt{\dfrac{\mu }{\mathsf{a}_{22}}}.
\end{equation}

\begin{figure}[H]
	\centering
	\includegraphics[scale=0.24,keepaspectratio]{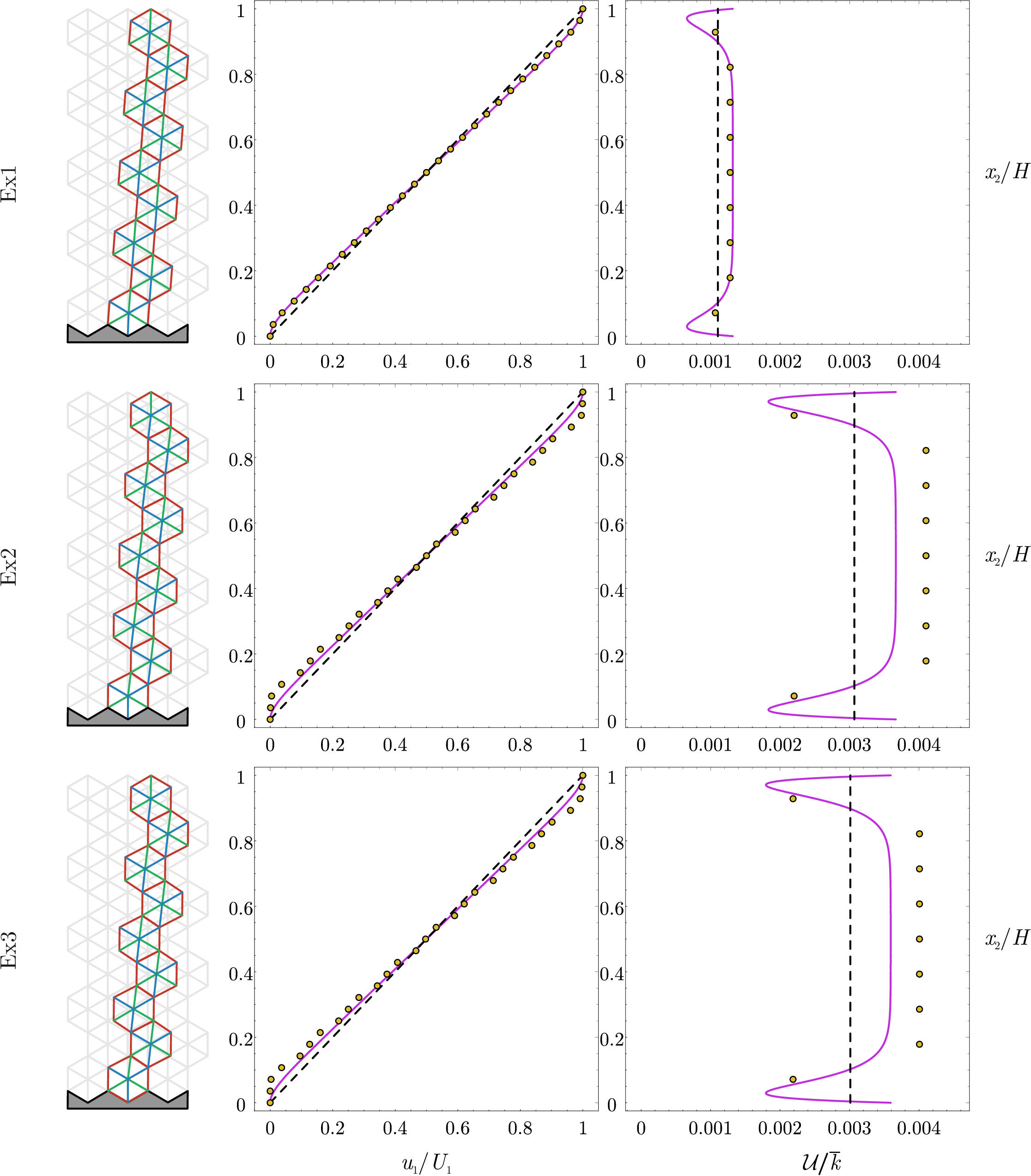}
	\caption{A simple shear strain is applied aligned parallel to the $x_1$--axis on a hexagonal lattice (left and reported as yellow dots on the right figures),  on a Cauchy (dashed line) and a second-gradient (purple line) equivalent solid. From left to right: Deformed configuration (superimposed to the undeformed configuration sketched gray), displacement field $u_1(x_2)$, and elastic strain energy density $\mathcal{U} (x_2)$ for the three cases labeled as Ex1 (upper part), Ex2 (central part), and Ex3 (lower part) as defined in Table \ref{tab:stiff_combo}. The enhancement in the modelling introduced through the second-gradient approximation are clearly visible.}
	\label{fig:Shear_x_A}
\end{figure}

Comparisons between the mechanical response of the lattice (solved exactly through Mathematica) and those of the equivalent Cauchy solid and of the equivalent $\mathsf{SGE}$ solid are reported in Fig. \ref{fig:Shear_x_A}, for the stiffness pairs labeled as Ex1 (upper part), Ex2 (central part), and Ex3 (lower part), see Table \ref{tab:stiff_combo}. In particular, the deformed configuration, the displacement $u_1(x_2)$, and the strain energy density $\mathcal{U}(x_2)$ are reported from left to right along the coordinate $x_2$. For the lattice, average values taken over each unit cell are reported as yellow dots, while the values evaluated within the equivalent Cauchy and $\mathsf{SGE}$ solids are reported as dashed and purple lines, respectively. The  improvement obtained when the lattice is modelled through the identified $\mathsf{SGE}$, with respect to the equivalent Cauchy material can be noted.

\subsubsection{Simple shear parallel to the $x_2$--axis}

In the case of a simple shear parallel to the $x_2$--axis, the governing equilibrium equations (\ref{eq:equilibrium_shear2}) become coupled. 
The solution of these equations can be obtained in a closed-form, but very complex, when the boundary conditions (\ref{BCsimpleshear}) are applied. The complication is related to the activation of the possible non-centrosymmetric effects. Therefore, this solution has been obtained with Mathematica, but is not reported.

\begin{figure}[H]
	\centering
	\includegraphics[scale=0.24,keepaspectratio]{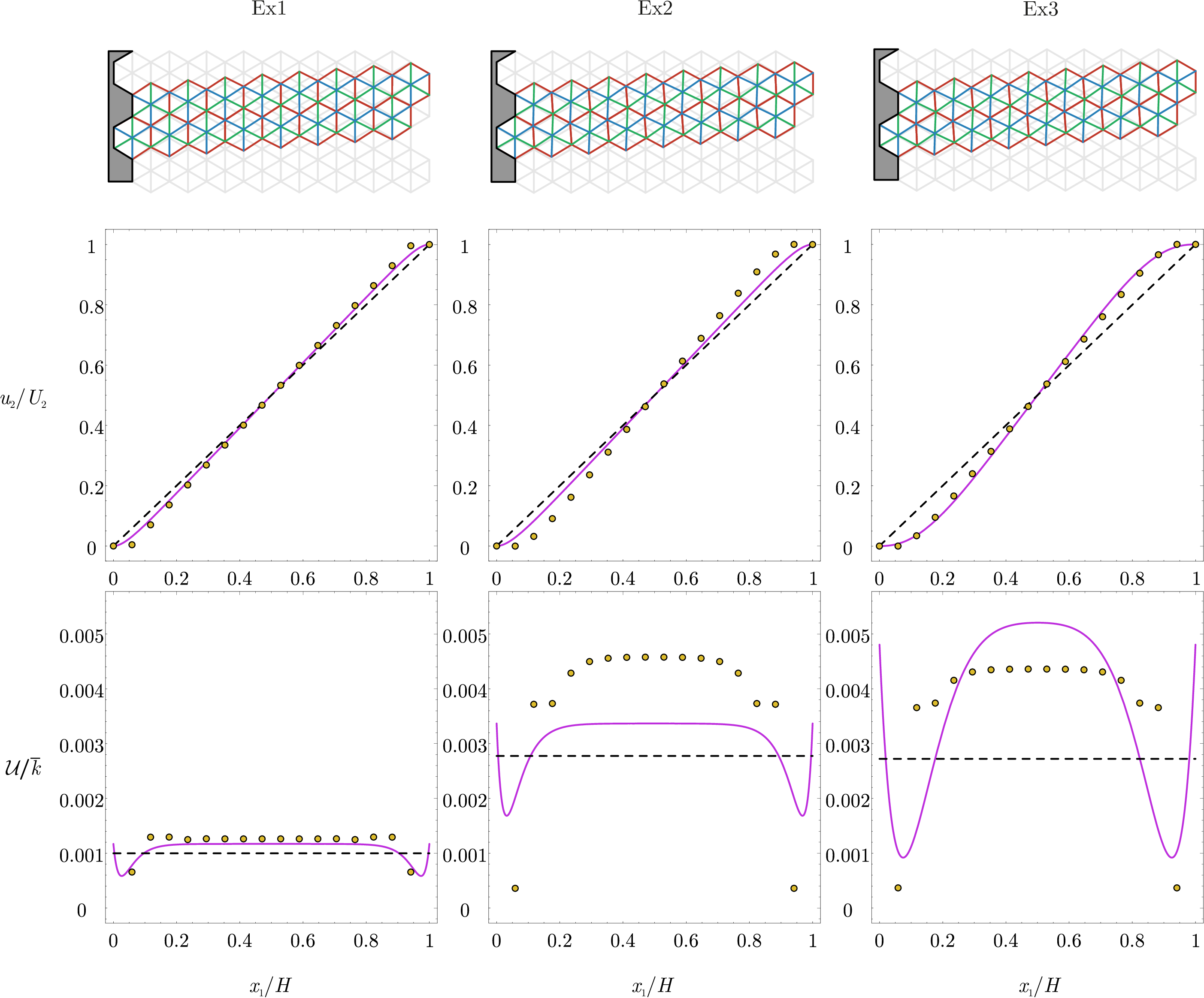}
	\caption{A simple shear strain is applied aligned parallel to the $x_2$--axis on a hexagonal lattice (upper part and reported as yellow dots on the lower figures),  on a Cauchy (dashed line) and a second-gradient (purple line) equivalent solid. From the upper to the lower part: deformed configuration (superimposed to the undeformed configuration sketched gray), displacement field $u_2(x_1)$, and strain energy density $\mathcal{U}(x_1)$ for the stiffness ratios labeled as Ex1, Ex2, and Ex3 in Table \ref{tab:stiff_combo}. }
	\label{fig:Shear_y_A}
\end{figure}

The mechanical responses of the lattice and of the two equivalent material models (Cauchy and $\mathsf{SGE}$) are reported in Fig. \ref{fig:Shear_y_A}, 
for stiffness ratios corresponding to the case labeled as Ex1, Ex2, and Ex3 in Table \ref{tab:stiff_combo} (from left to right in the figure). 
In the figure (from the upper part to the lower) 
the deformed configuration, the displacement field $u_2(x_1)$, and the strain energy density $\mathcal{U}(x_1)$ are shown.

\subsubsection{Validation of the second-gradient model for simple shear via energy mismatch}
The strain energy for the previously-reported simple shear problems has been evaluated with reference to a portion of the lattice, $\mathsf{U}_{\mathsf{lat}}$, and an equivalent portion of the identified materials (Cauchy, $\mathsf{U}_{\mathsf{Cau}}$, and second-gradient, $\mathsf{U}_{\mathsf{SGE}}$) shown in Figs. \ref{fig:Shear_x_A} and \ref{fig:Shear_y_A}. 
The difference between the strain energy in the lattice and the strain energy either of the Cauchy or of the second-gradient elastic materials are presented in Table~\ref{tab:ene_shear}, after a division by the strain energy in the lattice.\footnote{Note that the dimensionless difference in the total energy also coincides with the dimensionless difference in the resultant of the tractions, computed as $(\mathsf{R}_{IJ}-B \sigma_{IJ}^{\mathsf{Cau}})/\mathsf{R}_{IJ}$ for the equivalent Cauchy  and as $(\mathsf{R}_{IJ}-B P_I(\be_J))/\mathsf{R}_{IJ}$ for the equivalent $\mathsf{SGE}$. In these expressions, $\sigma_{IJ}^{\mathsf{Cau}}$ is the shear stress in the equivalent Cauchy solid and $\mathsf{R}_{IJ}$ is the component along the $x_J$--axis of the sum of the force in the bars of the lattice, cut with a plane with of normal $\be_I$ (being $I$ the direction of the simple shear, and $J\neq I$, with $\{I,J\}=\{1,2\}$).}  
The portion of the lattice considered for the evaluation of the elastic energy comprises a line of 9 unit cells for shear parallel to the $x_1$--axis and two lines of 8 unit cells 
for shear parallel to the $x_2$--axis, while a rectangular region has been used for the equivalent solids, respectively of dimensions 
$B=81\sqrt{3}\ell/28$ and $H=14\ell$, and   $B=48\ell/17$ and $H=17\sqrt{3}\ell/2$. The selected dimension of the rectangle ensures the correct area equivalence with the considered portion of lattice. 

The discrepancies in the strain energy reported in Table~\ref{tab:ene_shear} can be interpreted in a sense as the error made when the discrete lattice is treated as an equivalent elastic continuum. 

\begin{table}[H]
	\centering
	\caption{Errors in the strain energy matching between the lattice  and the  equivalent Cauchy and $\mathsf{SGE}$ solids for a simple shear parallel to the $x_1$--axis and parallel to the $x_2$--axis. The stiffness ratios labeled as Ex1, Ex2, and Ex3 in Table \ref{tab:stiff_combo} are considered. 	
	}
	\renewcommand{\arraystretch}{2.5}
\begin{tabular}{c|c|c|c|c|c|c|c|c|}
	\cline{2-5}
	& \multicolumn{2}{c|}{Simple shear parallel to $x_1$--axis} & \multicolumn{2}{c|}{Simple shear parallel to $x_2$--axis} \\ \cline{2-5} 
	&   $\dfrac{\mathsf{U}_{\mathsf{lat}}-\mathsf{U}_{\mathsf{Cau}}}{\mathsf{U}_{\mathsf{lat}}}$          &   $\dfrac{\mathsf{U}_{\mathsf{lat}}-\mathsf{U}_{\mathsf{SGE}}}{\mathsf{U}_{\mathsf{lat}}}$       &  $\dfrac{\mathsf{U}_{\mathsf{lat}}-\mathsf{U}_{\mathsf{Cau}}}{\mathsf{U}_{\mathsf{lat}}}$           &  $\dfrac{\mathsf{U}_{\mathsf{lat}}-\mathsf{U}_{\mathsf{SGE}}}{\mathsf{U}_{\mathsf{lat}}}$           \\ \hline
	\multicolumn{1}{|c|}{Ex1} &  10.46\%    &   2.14\%        &  16.26\%      &  9.38\%            \\ \hline
	\multicolumn{1}{|c|}{Ex2} & 16.58\%     &   8.83\%        &  26.74\%      &  19.3\%            \\ \hline
	\multicolumn{1}{|c|}{Ex3} & 16.4\%      &   8.69\%        &  25.63\%      &  1.14\%            \\ \hline
\end{tabular}
	\renewcommand{\arraystretch}{2.5}
	\label{tab:ene_shear}
\end{table}

The advantage in modelling the discrete lattice as an equivalent $\mathsf{SGE}$ material, instead of an equivalent Cauchy material, is evident from the Table, where the 
discrepancies in the strain energies are remarkably reduced using the second-gradient equivalent model, rather than the Cauchy solid. 

\subsection{Uniaxial strain problem}

The kinematical boundary conditions for a $\mathsf{SGE}$ material subject to a uniaxial strain  aligned parallel to the $x_I$--axis ($I=1,2$) are 
\begin{equation}
\label{BCuniax}
	\begin{array}{lllll}
	& u_I\left(x_I=0\right)=0,
	&u_I\left(x_I=H\right)=U_I,
	&u_{I,J}\left(x_I=0\right)=0,
	& u_{I,J}\left(x_I=H\right)=0,
	\\[3mm]
	&u_J\left(x_I=0\right)=0,&u_J\left(x_I=H\right)=0,
	&u_{J,J}\left(x_I=0\right)=0,
	&u_{J,J}\left(x_I=H\right)=0,
	\end{array}
	\qquad
	\mbox{for}\left\{
	\begin{array}{llll}
	I,J=1,2,\\
	J\neq I,
	\end{array}
	\right.
\end{equation} 
where $U_I$ is the imposed displacement aligned parallel to the $x_I$-direction and applied at $x_I=H$ (sketched with arrows on the right of Fig. \ref{fig:shearx1} for  the cases $I=1$ and $J=2$).
Because of the fact that the boundary conditions are independent of $x_J$ and the $\mathsf{SGE}$ material is homogeneous, the displacement fields are only dependent on  $x_I$,
\begin{equation}
u_I=u_I(x_I),\qquad
u_J=u_J(x_I),
\end{equation}
so that the equilibrium~(\ref{eq:eq_equa_gen}) in the case of uniaxial strain aligned parallel to the $x_1$--axis ($I=1$, $J=2$) reduces to Eqs.~(\ref{eq:equilibrium_shear2}) and Eqs.~(\ref{eq:traction_x2}), respectively, while for uniaxial strain aligned parallel the $x_2$--axis ($I=2$, $J=1$) to Eqs.~(\ref{eq:equilibrium_shear1}) and Eqs.~(\ref{eq:traction_x1}), respectively.

The equilibrium equations become uncoupled when the  uniaxial strain is imposed aligned parallel to the $x_2$--axis, while remain coupled when the uniaxial strain is aligned parallel the $x_1$--axis.

\subsubsection{Uniaxial strain parallel to the $x_2$--axis}

In terms of displacement field within the $\mathsf{SGE}$ solid, the solution of the equilibrium equations (\ref{eq:equilibrium_shear1}) when subjected to the boundary conditions (\ref{BCuniax}) yields the following expressions
\begin{equation}
u_{1}(x_2) = 0,
\quad
u_{2}(x_2) = \dfrac{ x_2 \cosh (\mathcal{Z}_{u}) + \dfrac{H}{2\mathcal{Z}_{u}} \left[\sinh \left(\mathcal{Z}_{u}\left(1 - \dfrac{2 x_2 }{H}\right)\right)-\sinh (\mathcal{Z}_{u})\right]}{ \cosh (\mathcal{Z}_{u}) - \dfrac{1}{\mathcal{Z}_{u}}\sinh(\mathcal{Z}_{u})} \dfrac{U_2}{H},
\end{equation}
while in terms of stress fields
\begin{equation}
	\begin{array}{ccccc}
	\sigma_{11}(x_2) = \dfrac{ \lambda  \left[ \cosh (\mathcal{Z}_{u})-\cosh \left(\mathcal{Z}_{u} \left(1-\dfrac{2 x_2}{H}\right)\right) \right] - \dfrac{2 \ell}{H} \mathsf{m} \mathcal{Z}_{u} \sinh \left(\mathcal{Z}_{u} \left(1-\dfrac{2 x_2}{H}\right)\right) }{ \cosh (\mathcal{Z}_{u})-\dfrac{1}{\mathcal{Z}_{u}}\sinh \left(\mathcal{Z}_{u} \right)} \dfrac{U_2}{H},
	\\[10mm]
	\sigma_{22}(x_2) = \dfrac{ \left(\lambda + 2\mu \right)  \left[ \cosh (\mathcal{Z}_{u})-\cosh \left(\mathcal{Z}_{u} \left(1-\dfrac{2 x_2}{H}\right)\right) \right] - \dfrac{2 \ell}{H} \mathsf{m} \mathcal{Z}_{u} \sinh \left(\mathcal{Z}_{u} \left(1-\dfrac{2 x_2}{H}\right)\right) }{ \cosh (\mathcal{Z}_{u})-\dfrac{1}{\mathcal{Z}_{u}}\sinh \left(\mathcal{Z}_{u} \right)} \dfrac{U_2}{H},
	\\[10mm]
	\end{array}
\end{equation}
and in terms of double stress fields
\begin{equation}
	\begin{array}{ccccc}
	\tau_{112}(x_2) =\ell \dfrac{ \mathsf{m}   \left[ \cosh (\mathcal{Z}_{u})-\cosh \left(\mathcal{Z}_{u} \left(1-\dfrac{2 x_2}{H}\right)\right) \right] + \dfrac{2 \ell}{H} \mathsf{a}_{34} \mathcal{Z}_{u} \sinh \left(\mathcal{Z}_{u} \left(1-\dfrac{2 x_2}{H}\right)\right) }{ \cosh (\mathcal{Z}_{u})-\dfrac{1}{\mathcal{Z}_{u}}\sinh \left(\mathcal{Z}_{u} \right)} \dfrac{U_2}{H}, \\[10mm]
	\tau_{222}(x_2) =\ell \dfrac{ -\mathsf{m}   \left[ \cosh (\mathcal{Z}_{u})-\cosh \left(\mathcal{Z}_{u} \left(1-\dfrac{2 x_2}{H}\right)\right) \right] + \dfrac{2 \ell}{H} \mathsf{a}_{44} \mathcal{Z}_{u} \sinh \left(\mathcal{Z}_{u} \left(1-\dfrac{2 x_2}{H}\right)\right) }{ \cosh (\mathcal{Z}_{u})-\dfrac{1}{\mathcal{Z}_{u}}\sinh \left(\mathcal{Z}_{u} \right)} \dfrac{U_2}{H}, \\[10mm]
	\tau_{211}(x_2) =\ell \dfrac{ \mathsf{m}   \left[ \cosh (\mathcal{Z}_{u})-\cosh \left(\mathcal{Z}_{u} \left(1-\dfrac{2 x_2}{H}\right)\right) \right] + \dfrac{2 \ell}{H} \mathsf{a}_{45} \mathcal{Z}_{u} \sinh \left(\mathcal{Z}_{u} \left(1-\dfrac{2 x_2}{H}\right)\right) }{ \cosh (\mathcal{Z}_{u})-\dfrac{1}{\mathcal{Z}_{u}}\sinh \left(\mathcal{Z}_{u} \right)} \dfrac{U_2}{H}, \\[10mm]
	\end{array}
\end{equation}
showing that, although the governing equations (\ref{eq:equilibrium_shear1}) are decoupled, the stress and double-stress fields depend on the non-centrosymmetry constitutive parameter.

The components of the resultant traction vector associated with the normal $\bn=\be_2$ are found to be independent of $x_2$ in the form

\begin{equation}
P_{1}\left(\be_{2}\right) = 0, \qquad
P_{2}\left(\be_{2}\right) = \dfrac{\lambda + 2\mu}{1 - \dfrac{1}{\mathcal{Z}_{u}}\tanh\mathcal{Z}_{u}}\dfrac{U_2}{H},
\end{equation}
and the three contributions $\mathcal{U}_{\boldsymbol{\mathsf{C}}}$, $\mathcal{U}_{\boldsymbol{\mathsf{M}}}$, and $\mathcal{U}_{\boldsymbol{\mathsf{A}}}$ to the strain energy density are given by
\begin{equation}
\begin{array}{ccccc}
\mathcal{U}_{\boldsymbol{\mathsf{C}}}(x_2)=
\dfrac{\left(\lambda + 2\mu\right)\left[ \cosh\left( \mathcal{Z}_{u} \right ) - \cosh\left( \mathcal{Z}_{u}\left(1-\dfrac{2x_2}{H} \right ) \right ) \right ]^2}{2\left(\cosh\left( \mathcal{Z}_{u} \right ) - \dfrac{1}{\mathcal{Z}_{u}} \sinh\left( \mathcal{Z}_{u} \right )\right)^2},
\\[10mm]
\mathcal{U}_{\boldsymbol{\mathsf{M}}}(x_2)= \dfrac{2 \ell}{H}\dfrac{ \mathsf{m} \mathcal{Z}_{u} \sinh \left(\mathcal{Z}_{u} \left(1-\dfrac{2 x_2}{H}\right)\right) \left[\cosh \left(\mathcal{Z}_{u} \left(1-\dfrac{2x_2}{H}\right)\right)-\cosh (\mathcal{Z}_{u})\right]}{\left(\cosh (\mathcal{Z}_{u})-\dfrac{\sinh (\mathcal{Z}_{u})}{\mathcal{Z}_{u}}\right)^2} \dfrac{U_2^2}{H^2}, 
\\[10mm]
\mathcal{U}_{\boldsymbol{\mathsf{A}}}(x_2)=
\dfrac{ \left(\lambda + 2\mu \right) \sinh^2 \left( \mathcal{Z}_{u}\left(1-\dfrac{2x_2}{H} \right ) \right )}{2\left(\cosh\left( \mathcal{Z}_{u} \right ) - \dfrac{1}{\mathcal{Z}_{u}} \sinh\left( \mathcal{Z}_{u} \right )\right)^2},
\end{array}
\end{equation}
where
\begin{equation}
\mathcal{Z}_{u}=\dfrac{H }{2 \ell}\sqrt{\dfrac{\lambda + 2\mu }{\mathsf{a}_{44}}}.
\end{equation}

\begin{figure}[H]
	\centering
	\includegraphics[scale=0.24,keepaspectratio]{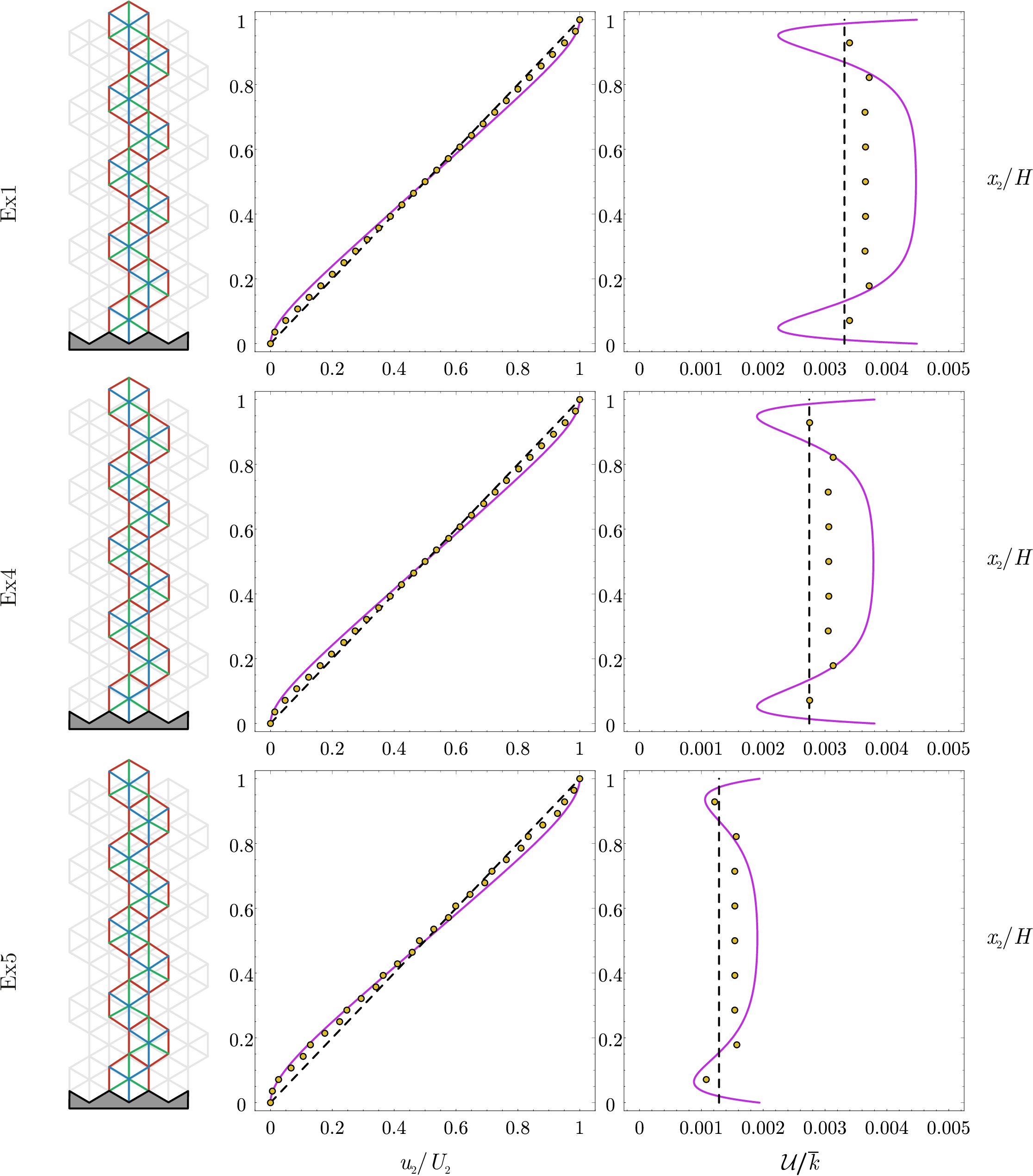}
	\caption{A uniaxial strain is applied aligned parallel to the $x_2$--axis on a hexagonal lattice (left and reported as yellow dots on the right figures),  on a Cauchy (dashed line) and a second-gradient (purple line) equivalent solid. From left to right: Deformed configuration (superimposed to the undeformed configuration sketched gray), displacement field $u_2(x_2)$, and elastic strain energy density $\mathcal{U} (x_2)$ for the three cases labeled as Ex1 (upper part), Ex4 (central part), and Ex5 (lower part),  as defined in Table \ref{tab:stiff_combo}.}
	\label{fig:Uniaxial_y_A}
\end{figure}

Comparisons between the mechanical response of the lattice (solved exactly through Mathematica) and those of the equivalent Cauchy solid and of the equivalent $\mathsf{SGE}$ solid are reported in Fig. \ref{fig:Uniaxial_y_A}, for the stiffness pairs labeled as Ex1 (upper part), Ex4 (central part), and Ex5 (lower part), in Table \ref{tab:stiff_combo}. In particular, the deformed configuration, the displacement $u_2(x_2)$, and the strain energy density $\mathcal{U}(x_2)$ are reported from left to right along the coordinate $x_2$.
For the lattice, average values taken over each unit cell are reported as yellow dots, while the values evaluated within the equivalent Cauchy and $\mathsf{SGE}$ solids are reported as dashed and purple lines, respectively.

It is highlighted that, for Ex5, the strain energy density of the $\mathsf{SGE}$ solid is not symmetric with respect to the horizontal line drawn at mid point of the height $H$.
The asymmetry is due to the non-centrosymmetric component $\mathcal{U}_{\mathsf{M}}$ of the strain energy density, a component which results skew-symmetric.
The  improvement obtained when the lattice is modelled through the identified $\mathsf{SGE}$, with respect to the equivalent Cauchy material, can be noted.

\subsubsection{Uniaxial strain parallel to the $x_1$--axis}
\label{stokaz}

In the case of a uniaxial strain parallel to the $x_1$--axis, the governing equilibrium equations (\ref{eq:equilibrium_shear2}) become coupled. 
The solution of these equations can be obtained in a closed, but very complicated, form when the boundary conditions (\ref{BCuniax}) are applied. The complexity is related to the non-centrosymmetric effect. Therefore, the closed-form solution obtained with Mathematica is not reported.

\begin{figure}[H]
	\centering
	\includegraphics[scale=0.24,keepaspectratio]{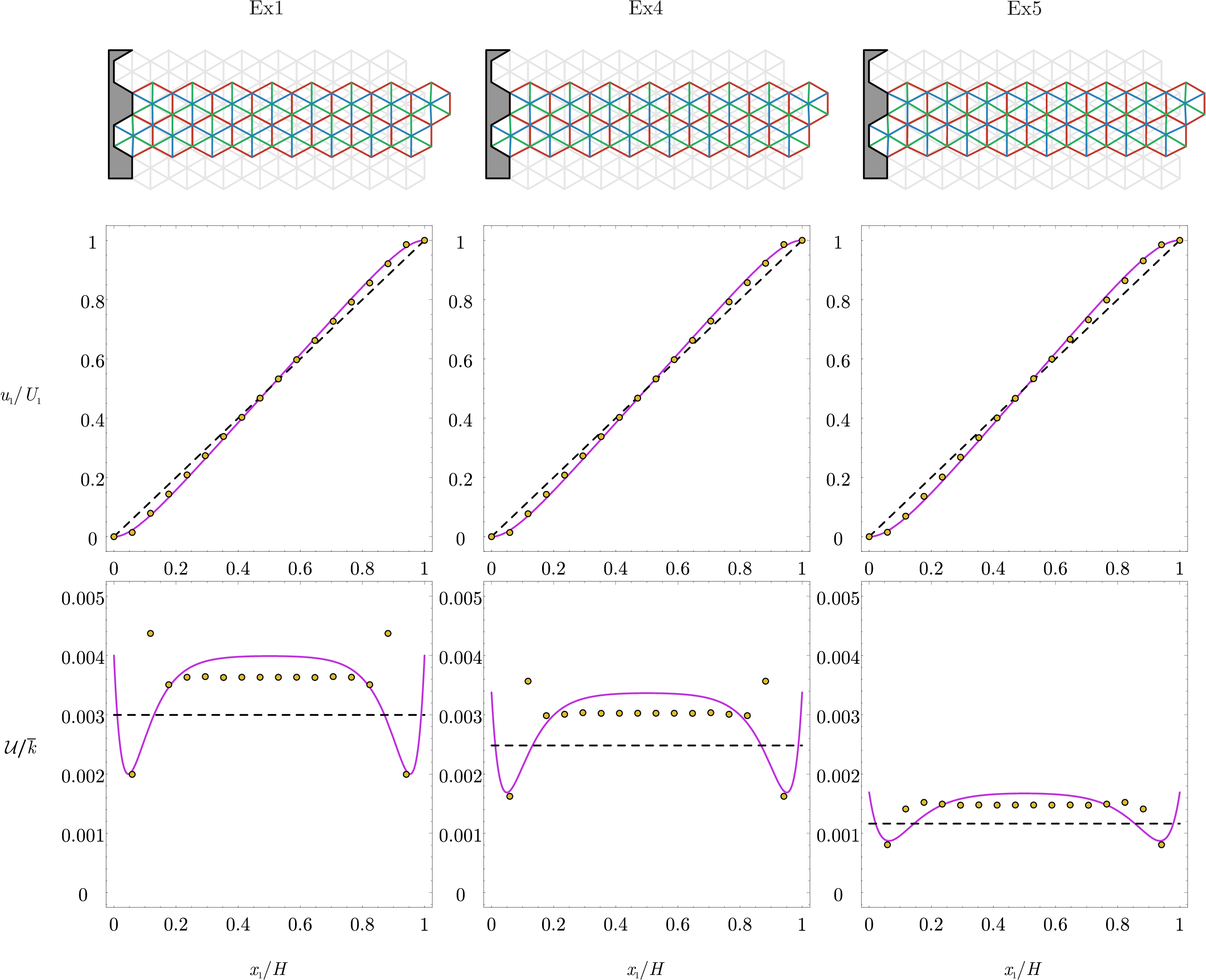}
	\caption{A uniaxial strain aligned parallel to the $x_1$--axis on a hexagonal lattice (upper part and reported as yellow dots on the lower figures), on a Cauchy (dashed line) and a second-gradient (purple line) equivalent solid. From the upper to the lower part: deformed configuration (superimposed to the undeformed configuration sketched gray), displacement field $u_1(x_1)$, and strain energy density $\mathcal{U}(x_1)$ for the stiffness ratios labeled as Ex1, Ex4, and Ex5 in Table \ref{tab:stiff_combo}.}
	\label{fig:Uniaxial_x_A}
\end{figure}

The mechanical responses of the lattice and of the two equivalent solids (Cauchy and second-gradient) are reported in Fig. \ref{fig:Uniaxial_x_A}, for stiffness ratios corresponding to the case labeled as Ex1, Ex4, and Ex5 in Table \ref{tab:stiff_combo} (found from left to right in the figure). 
The deformed configuration, the displacement field $u_1(x_1)$, and the strain energy density $\mathcal{U}(x_1)$ are shown in the figure (from the upper part to the lower).

\subsubsection{Validation of the second-gradient model for uniaxial strain via energy mismatch}

The strain energy for the previously-reported uniaxial strain problems has been evaluated with reference to a portion of the lattice, $\mathsf{U}_{\mathsf{lat}}$, and an equivalent portion of the identified materials (Cauchy, $\mathsf{U}_{\mathsf{Cau}}$, and second-gradient, $\mathsf{U}_{\mathsf{SGE}}$) shown in Figs. \ref{fig:Uniaxial_y_A} and \ref{fig:Uniaxial_x_A}. 
The difference between the strain energy in the lattice and the strain energy either of the Cauchy or of the second-gradient elastic materials are presented in Table~\ref{tab:ene_uniax}, after a division by the strain energy in the lattice.\footnote{As in the simple shear problem, for the uniaxial strain problems the dimensionless difference in the resultant of the tractions coincides with that in the energy.}
The portion of the lattice considered for the evaluation of the elastic energy comprises a line of 9 unit cells for uniaxial strain parallel to the $x_2$--axis and two lines of 8 unit cells for uniaxial strain parallel to the $x_1$--axis, while a rectangular region has been used for the equivalent solids, respectively of dimensions $B=81\sqrt{3}\ell/28$ and $H=14\ell$, and  $B=48\ell/17$ and $H=17\sqrt{3}\ell/2$. The selected dimension of the rectangle ensures the correct equivalence with the considered portion of lattice. 

The discrepancies in the strain energy reported in Table~\ref{tab:ene_uniax} can be interpreted in a sense as the error made when the discrete lattice is treated as an equivalent elastic continuum. 

\begin{table}[H]
	\centering
	\caption{Errors in the strain energy matching between the lattice  and the  equivalent Cauchy and $\mathsf{SGE}$ solids for a uniaxial strain parallel to the $x_1$--axis and parallel to the $x_2$--axis. The stiffness ratios labeled as Ex1, Ex4, and Ex5 in Table \ref{tab:stiff_combo} are considered.}
	\renewcommand{\arraystretch}{2.5}
	\begin{tabular}{c|c|c|c|c|c|c|c|c|}
		\cline{2-5}
		& \multicolumn{2}{c|}{Uniaxial strain parallel to $x_1$--axis} & \multicolumn{2}{c|}{Uniaxial strain parallel to  $x_2$--axis} \\ \cline{2-5} 
		&   $\dfrac{\mathsf{U}_{\mathsf{lat}}-\mathsf{U}_{\mathsf{Cau}}}{\mathsf{U}_{\mathsf{lat}}}$          &   $\dfrac{\mathsf{U}_{\mathsf{lat}}-\mathsf{U}_{\mathsf{SGE}}}{\mathsf{U}_{\mathsf{lat}}}$    &  $\dfrac{\mathsf{U}_{\mathsf{lat}}-\mathsf{U}_{\mathsf{Cau}}}{\mathsf{U}_{\mathsf{lat}}}$           &  $\dfrac{\mathsf{U}_{\mathsf{lat}}-\mathsf{U}_{\mathsf{SGE}}}{\mathsf{U}_{\mathsf{lat}}}$        \\ \hline
		\multicolumn{1}{|c|}{Ex1} &  8.16\%    &   -6.83\%             &  14.52\%      &  -1.24\%      \\ \hline
		\multicolumn{1}{|c|}{Ex4} & 8.67\%     &   -7.47\%             &  14.75\%      &  0.57\%       \\ \hline
		\multicolumn{1}{|c|}{Ex5} & 11.93\%    &   -8.23\%             &  16.5\%       &  -0.72\%     \\ \hline
	\end{tabular}
	\renewcommand{\arraystretch}{2.5}
	\label{tab:ene_uniax}
\end{table}

The negative values reported in Table \ref{tab:ene_uniax} for the discrepancy in the strain energy are referred to situations where the second-gradient equivalent 
solid is stiffer than the lattice. While the advantage in modelling the lattice as an equivalent second-gradient continuum is not particularly 
evident from the case of uniaxial strain parallel to the $x_1$--axis, the advantage becomes clear when the uniaxial strain becomes parallel to the $x_2$--axis.

\subsection{Non-centrosymmetry in the lattice and in the SGE material}

\begin{figure}[H]
	\centering
	\includegraphics[width=0.94\textwidth,keepaspectratio]{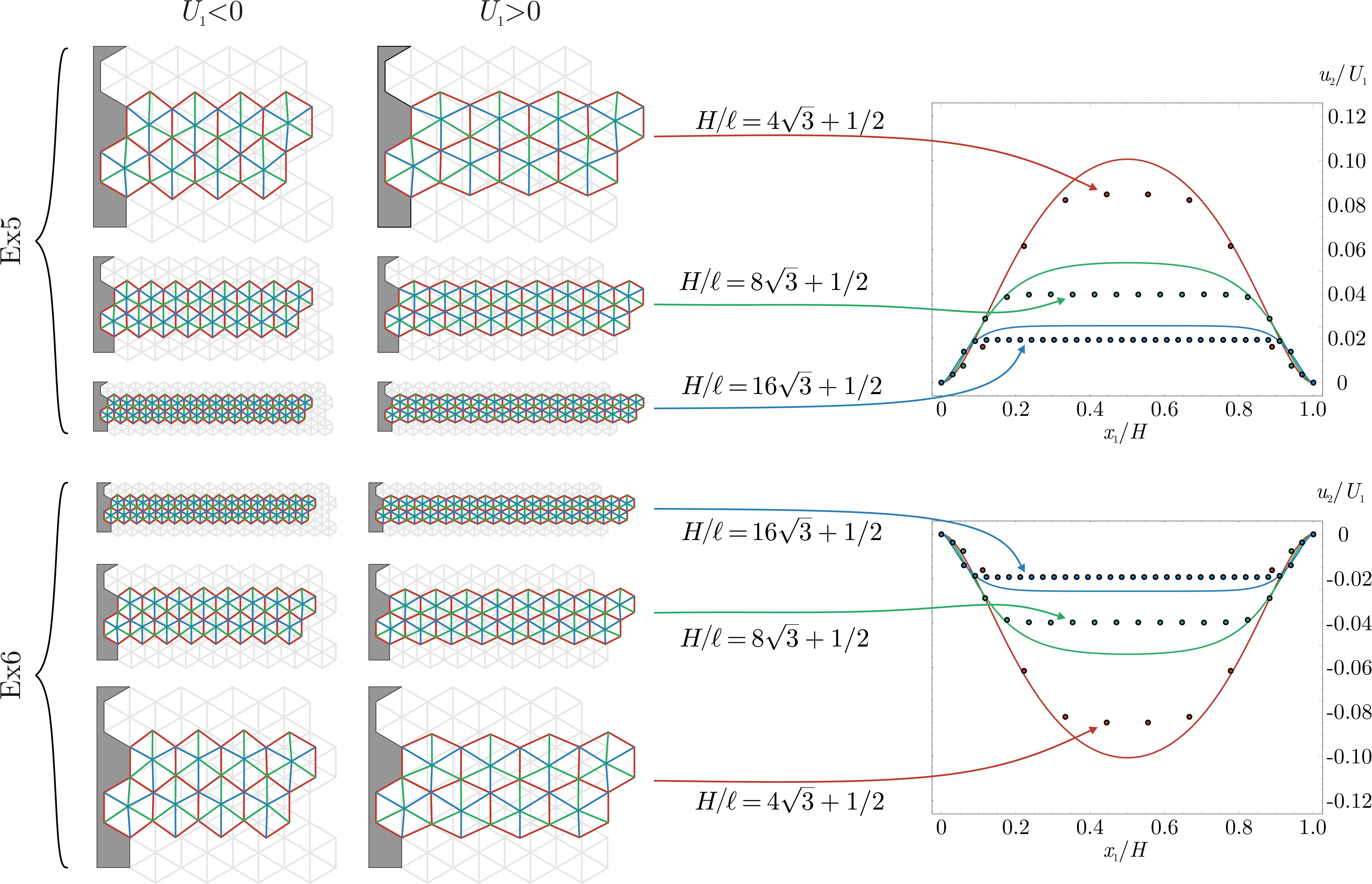}
	\caption{Deformed configurations (superimposed to the undeformed configuration sketched gray) and dimensionless transverse displacement $u_2(x_1)/U_1$ for the lattice and the corresponding equivalent $\mathsf{SGE}$ solid, in the cases labeled as Ex5 and Ex6 in Table \ref{tab:stiff_combo} and subject to compressive ($U_1<0$) and tensile ($U_1>0$) uniaxial strain aligned parallel to the $x_1$--axis. Different geometries are reported, $H/\ell=\{4\sqrt{3}+1/2, 8\sqrt{3}+1/2, 16\sqrt{3}+1/2\}$.
	Note that the transverse displacement would be null for a Cauchy equivalent material, thus evidencing the superiority of the second-gradient model.
		}
	\label{fig:Uniaxial_x_NCS}
\end{figure}

The purpose of this section is to provide a clear evidence of the non-centrosymmetric effect in the lattice structure and its proper prediction through the identified equivalent $\mathsf{SGE}$ solid.
In fact, the non-centrosymmetry of a material can be interpreted as a coupling between strains and curvatures, so that if a non-centrosymmetric solid is subjected to strain, curvature is generated and vice-versa.
	A basic example for which the lattice and, in agreement, the equivalent $\mathsf{SGE}$ solid show non-centrosymmetric behaviour is the case of uniaxial strain aligned parallel to the $x_1$--axis (Section \ref{stokaz}), whenever the three bars' stiffnesses $\overline{k}$, $\widehat{k}$, and $\widetilde{k}$ do not satisfy Eq.~(\ref{eq:constrain_CS}).
	In this case, the lattice structure displays a transverse displacement $u_2(x_1)$, which is simply not captured by the equivalent Cauchy solid, but is correctly modelled with the second-gradient equivalent continuum, 
		as shown in Fig. \ref{fig:Uniaxial_x_NCS}. 
		
	Fig. \ref{fig:Uniaxial_x_NCS} reports results for lattices and corresponding $\mathsf{SGE}$ solids characterized by the stiffness ratios labeled as Ex5 and Ex6 in Table \ref{tab:stiff_combo} and subject to  compressive ($U_1<0$) and tensile ($U_1>0$) uniaxial strain, both applied to three different hexagonal geometries ($H/\ell=\{4\sqrt{3}+1/2, 8\sqrt{3}+1/2, 16\sqrt{3}+1/2\}$). 
It is possible to observe from the figure that the transverse displacement field $u_2(x_1)$ takes a different sign 
in the upper and lower parts of Fig. \ref{fig:Uniaxial_x_NCS}. This is a consequence of the fact that the equivalent constitutive parameter $\boldsymbol{\mathsf{m}}$ keeps the same absolute value, but changes its sign, when two the non-perimeter springs are permuted.
Note that, due to the normalization, the diagrams $u_2(x_1)/U_1$ are insensitive to the sign of the imposed displacement $U_1$.

\section{Conclusions}

A class of second-gradient elastic materials has been derived from  a \lq condensed' form of the constitutive tensors,  
 identified in Part I of this study to be energetically equivalent to a hexagonal (planar and infinite) truss structure made up of three orders of different elastic bars. 
It is shown that the identified homogeneous material displays mechanical properties of: (i.) non-locality, (ii.) non-centrosymmetry, and (iii.) anisotropy (although the local behaviour is isotropic, in consequence of the hexagonal geometry, local, and centrosymmetric).
The capability of the presented model is assessed through comparisons of the mechanical response under simple shear and uniaxial strain for different lattice structures and their equivalent material counterparts, for which analytical solutions have been derived. The performed comparisons provide a quantitative validation of the present identification technique, thus confirming its applicative potentiality in the advanced design of microstructured solids.

\paragraph{Acknowledgements.} GR, DV, FDC gratefully
acknowledge financial support from the grant ERC Advanced Grant \lq
Instabilities and nonlocal multiscale modelling of materials' ERC-2013-ADG-340561-INSTABILITIES. DB gratefully
acknowledges financial support from PRIN 2015 \lq Multi-scale mechanical models for the design and optimization of micro-structured smart materials and metamaterials' 2015LYYXA8-006.
\bibliographystyle{plain}
\bibliography{Biblio2}
\end{document}